 \documentclass[final,5p,times,twocolumn]{elsarticle}
 \usepackage[normalem]{ulem}  
\usepackage{xcolor}          
\usepackage{amssymb}
\usepackage{lipsum}
\usepackage{algorithm}
\usepackage{algpseudocode}
\usepackage{amsmath}
\usepackage{svg}
\usepackage{stfloats}
\usepackage{hyperref}
\usepackage{multirow}
\usepackage{listings}
\usepackage[pagewise]{lineno}
\usepackage{zref-user}

\NewDocumentCommand{\codeword}{v}{%
\texttt{\textcolor{blue}{#1}}%
}

\journal{Nuclear Physics A}

\begin{document}
\begin{frontmatter}



\title{Small-Cell-Based Fast Active Learning of Machine Learning Interatomic Potentials}


\author[inst1]{Zijian Meng\corref{cor1}}
\ead{17zjm1@queensu.ca; contact@richardzjm.com}
\fntext[fn1]{Footnote for the first author.}
\cortext[cor1]{Corresponding author.}

\author[inst1]{Hao Sun}
\ead{hs126@queensu.ca}

\author[inst2]{Edmanuel Torres}
\ead{edmanuel.torres@cnl.ca}

\author[inst2]{Christopher Maxwell}
\ead{christopher.maxwell@cnl.ca}

\author[inst3]{Ryan Eric Grant}
\ead{ryan.grant@queensu.ca}

\author[inst1]{Laurent Karim Béland}
\ead{laurent.beland@queensu.ca}

\affiliation[inst1]{organization={Department of Mechanical and Materials Engineering, Queen's University},
                    city={Kingston}, 
                    state={Ontario},
                    country={Canada, K7L 3N6}}

\affiliation[inst3]{organization={Department of Electrical and Computer Engineering, Queen's University},
                    city={Kingston}, 
                    state={Ontario},
                    country={Canada, K7L 3N6}}

\affiliation[inst2]{organization={Canadian Nuclear Laboratories},
                    city={Chalk River}, 
                    state={Ontario},
                    country={Canada}}

\begin{abstract}

Machine learning interatomic potentials (MLIPs) are often trained with on-the-fly active learning, where sampled configurations from atomistic simulations are added to the training set. However, this approach is limited by the high computational cost of \textit{ab initio} calculations for large systems. Recent works have shown that MLIPs trained on small cells (1–8 atoms) rival the accuracy of large-cell models (100s of atoms) at far lower computational cost. Herein, we refer to these as small-cell and large-cell training, respectively. In this work, we iterate on earlier small-cell training approaches and characterize our resultant small-cell protocol. Potassium and sodium-potassium systems were studied: the former, a simpler system benchmarked in detail; the latter, a more complex binary system for further validation. Our small-cell training approach achieves up to two orders of magnitude of cost savings compared to large-cell (54-atom) training, with some training runs requiring fewer than 120 core-hours. Static and thermodynamic properties predicted using the MLIPs were evaluated, with small-cell training in both systems yielding strong \textit{ab initio} agreement. Small cells appear to encode the necessary information to model complex large-scale phenomena—solid-liquid interfaces, critical exponents, diverse concentrations—even when the training cells themselves are too small to accommodate these phenomena. Based on these tests, we provide analysis and recommendations.

\end{abstract}




\end{frontmatter}




\section{Introduction}
\label{introduction}

Machine learning interatomic potentials (MLIPs) are an emerging class of functional forms for modeling atomic interactions~\cite{deringer2019machine,mishin2021machine,mueller2020machine}, including in molecular dynamics (MD) simulations. MLIPs bridge the accuracy-cost gap between traditional semi-analytical interatomic potentials and \textit{ab initio} methods. Usually trained on \textit{ab initio} data—commonly density functional theory (DFT)—MLIPs can be systematically improved to target a desired accuracy-cost trade-off. Few physics-based assumptions are made about the form of the final potential, aside from imposing symmetries. Instead, MLIPs rely on fitting general and transferable mathematical functions (or processes) to the \textit{ab initio} dataset. MLIPs generally comprise a descriptor that extracts features from local atomic environments and a regressor that interprets these features. Common descriptors include Behler symmetry functions~\cite{behler2007generalized}, smooth overlap of atomic orbitals (SOAP)~\cite{bartok2013representing}, bispectrum coefficient descriptors~\cite{bartok2010gaussian}, and moment tensors~\cite{shapeev2016moment}. Common regressors include neural networks, linear regression, and Gaussian process regression.

MLIPs are fitted to training sets of configurations of atoms, usually either curated using domain knowledge (supervised learning) or obtained using active learning. In active learning, candidate configurations are first sampled, then an uncertainty quantification (UQ) metric is calculated to select which configurations are unlikely to be correctly described by the current MLIP. These configurations are evaluated with \textit{ab initio} methods, and the results are added to the training set~\cite{settles2009active, li2024local}. In on-the-fly active learning, a preliminary MLIP is employed to run an atomistic simulation, generating a candidate configuration at each step. Based on the chosen UQ metric, some candidate configurations are added to the training set, after which the MLIP is retrained. Following this process iteratively—repeating the simulations using the re-trained MLIP—leads to improvement of the MLIP's description of the simulation~\cite{jinnouchi2020fly, verdi2021thermal, podryabinkin2017active}. Various UQ strategies exist, such as query-by-committee, Gaussian uncertainty, clustering, and furthest point sampling~\cite{zhu2023fast, li2024local,huan2017universal}. Importantly, UQ itself must avoid expensive \textit{ab initio} calculations. In most strategies, if the level of uncertainty exceeds a predefined threshold, the configuration is selected, evaluated by \textit{ab initio} methods, and added to the training set.

\subsection{Active Learning Machine Learning Potentials}

Early active learning research utilized Bayesian query-by-committee. Frederiksen~\textit{et al.} suggested that comparisons between potentials could be used for training set optimization~\cite{frederiksen2004bayesian}. Behler then expanded upon query-by-committee, proposing its use for UQ and on-the-fly training~\cite{behler2014representing}. Smith~\textit{et al.} applied query-by-committee active learning to biomolecules with their ANI-1 neural network potential~\cite{smith2017ani}. Using UQ to reduce their initial training set size and automatically sampling chemical spaces from existing datasets, their final potential outperforms their previous ANI-1 potential with only 25\% of the training set size.

Jinnouchi~\textit{et al.} used an on-the-fly MD algorithm, leveraging the Gaussian approximation potential to predict the melting points of several metals~\cite{jinnouchi2019fly}. Their method relies on prior sampling history and Bayesian linear regression UQ to decide whether or not to perform \textit{ab initio} calculations and store them. If the level of uncertainty or the number of stored configurations surpasses a threshold, the training set is updated and the potential is retrained.  

Podryabinkin~\textit{et al.} introduced a MaxVol-based UQ metric targeting linear regression models, including the Moment Tensor Potential (MTP)~\cite{podryabinkin2017active,goreinov2010find}. The MaxVol strategy, as implemented in the MLIP-2 software package, employed two-UQ-threshold bootstrapping~\cite{novikov2020mlip}. Configurations whose uncertainty exceeded a selection threshold were stored for later evaluation, while simulations visiting configurations with a high level of uncertainty were terminated immediately. Once the MD runs were terminated, stored configurations were sent to DFT code, and the results were added to the training set. The potential was then retrained, and the simulation restarted. Later, Podryabinkin~\textit{et al.} introduced a neighborhood-based MaxVol UQ in MLIP-3~\cite{podryabinkin2017active, podryabinkin2023mlip}, which extracts clusters of atoms from large atomic configurations for use in non-periodic DFT calculations—thus facilitating on-the-fly learning with larger cells. Other methods to prepare DFT cells during on-the-fly learning include the QM/ML approach by Grigorev~\textit{et al.}~\cite{grigorev2021synergistic}.

Due to the superlinear cost of DFT, which scales with the atom (electron) count, on-the-fly learning cannot be trivially applied to simulation boxes containing thousands or millions of atoms. This is an issue as many problems of scientific interest require simulation boxes of such size. Additionally, active learning schemes typically involve variable-core compute workloads, which are impractical in many high-performance computing environments. Extracting clusters of atoms as done in MLIP-3~\cite{podryabinkin2023mlip} can help reduce costs. However, the extracted clusters will need to be at least as large as the outer cutoff radius of the potential~\cite{podryabinkin2023mlip}—often even larger, as one needs to ensure that there is no significant electronic interaction between the surface of the cluster and the region of interest.

\subsection{Small-Cell Training}
To mitigate superlinear DFT compute scaling, one can train on smaller cells. This approach is commonly used for modeling perfect, periodic crystal behavior (e.g., equation of state or elastic response) but is typically augmented with larger cells to capture phenomena like surfaces, defects, and phase transitions. Seldom are MLIPs trained solely or even primarily on small-cell configurations. 

To our knowledge, we observe two main approaches to small-cell-centric training: one using direct (possibly random) sampling of configuration space, and the other, sampling along curated trajectories using active learning.

Pickard describes one such small-cell protocol in which \textit{ab initio} random structure searching is used to rapidly construct a shallow neural network potential for several systems, including boron~\cite{pickard2022ephemeral}. Training on only 8-atom cells, his potential was able to identify the 12-atom icosahedral $\alpha$-boron and the complex 28-atom $\gamma$-boron structure. 

Pozdynakov \textit{et al.} trained their GTTP model for aluminum using structures produced by a symmetric random structure generator from evolutionary algorithm USPEX\cite{PhysRevB.107.125160,lyakhov2013new}. Although trained with only 8-atom cells, their MLIP produced satisfactory accuracy when tested against large structures. Similarly, Poul \textit{et al.} used randomized variations on \textit{ab initio} relaxation trajectories of 1 to 10 atoms in various crystal structures generated through RandSpg to train a magnesium MTP~\cite{poul2023systematic,avery2017randspg}. 

Meziere~\textit{et al.} used small cells to train a zirconium hydride~\cite{meziere2023accelerating} MTP. Active learning was combined with geometric relaxation of supercells of increasing size. Their MTP was ultimately trained using a set where 95\% of members contained 7 or fewer atoms but still successfully described the Zr-H convex hull. Meziere \textit{et al.} then showed that the resultant training set could be expanded with on-the-fly MD of a single $\alpha$ phase heating and a single $\beta$ phase heating simulation whose cell sizes were iteratively increased. Using 8 atoms per cell or less, their final potential captured the $\alpha$-$\beta$ phase transition with similar accuracy to on-the-fly training on cells containing 48 or 54 atoms. A 17.5 times CPU speedup versus large cells was obtained. 

Luo~\textit{et al.} later applied small-cell training on Zr MTPs. They also used small-cell simulations of increasing size~\cite{luo2023set}. A set of small-cell (less than 15-atom) MD active learning simulations was first performed; medium-cell simulations involving up to 38 atoms followed. Then, large-cell active learning MD simulations containing up to 512 atoms were performed, with no large-cell configurations added to the training set. 45\% of the final training set was comprised of small-cell configurations.

Small-cell training was applied by Sun \textit{et al.} to train a NaCl MTP~\cite{sun2024interatomic}. The training first considered primitive and unit cells before transitioning to liquid simulations of up to 34 atoms. The resultant MTP described pure Na and NaCl in both solid and liquid forms, as well as gaseous Cl. Only 449 learnable parameters were trained on 609 configurations, 55\% of which were unit or primitive cells.

Our proposed approach adopts aspects of both of the two existing approaches to small-cell training, randomly sampling a range of small-cell MD trajectories in parallel. This augments the stability and guided training of earlier active learning small-cell methods with the diverse exploration range of random sampling small-cell methods to train a potential for various test environments.

Herein, we investigate the advantages and limitations of small-cell training in two case studies, using the MTP formalism as implemented in the MLIP-3 software package \cite{podryabinkin2023mlip}. First, we characterize small-cell training in greater depth than previous works using a simple monoatomic system—potassium, varying parameters such as ML model parameters, cell sizes, and UQ methodology. Second, we expand our approach to a binary system, sodium-potassium alloy, and validate small-cell training in a more complex system. Many of our findings are generalizable to other local MLIP formalisms, as long as UQ and active learning are available in MD simulations with periodic boundaries. 

We begin with an introduction to the MTP-MaxVol algorithm and propose our systematic small-cell training protocol based on a parallel MD query strategy, implemented as a fully automated workflow. Our computational cost benchmarks are presented, and we test the MTPs against DFT and experimental data. Finally, we offer recommendations for future small-cell training work. 

\section{Methodology}
We begin with an abbreviated examination of the active learning strategy for training MTPs described in Refs.~\cite{podryabinkin2017active, goreinov2010find,novikov2020mlip,podryabinkin2023mlip}, implemented in the MLIP-2 and MLIP-3 software packages. Afterward, we describe a systematic small-cell active learning protocol. For details of the MTP framework and descriptors, we refer the reader to the excellent explanations provided in Refs.~\cite{shapeev2016moment,podryabinkin2017active,novikov2020mlip,podryabinkin2023mlip}. 

\subsection{Maxvol-Based Active Learning}
The Maxvol-based active learning is motivated by the D-Optimality criterion. An active set is maintained, which represents the convex hull of the configurational space of the training set as defined by the MTP descriptors. Candidate samples each have an associated extrapolation grade, which is defined by how much it increases the hull's volume. This extrapolation grade is the uncertainty metric.

Each sample, $X$, is either a configuration, $\textrm{cfg}$, or a neighborhood, $\textrm{nbh}$, depending on the uncertainty quantification mode used. For configurations, the predicted energy of the configuration, $E_{\textrm{mtp}}$, is linearized; for neighborhood mode, the predicted energy of the corresponding neighborhood, $V_{\textrm{mtp}}$, is linearized.  For $n$ learnable parameters in a vector $\theta$, we assume $\theta$ to be near its optimal values, $\overline{\theta}$. Then, for a sample, $X$, the linearized predicted energy relative to the $i$\textsuperscript{th} parameter is given as:

\begin{equation}
    b_i(\textrm{X}) = \frac{\partial}{\partial\overline{\theta_i}} E_{\textrm{mtp}}(\textrm{cfg},\overline{\theta})
\end{equation} for configurations or
\begin{equation}
    b_i(\textrm{X}) = \frac{\partial}{\partial\overline{\theta_i}} V_{\textrm{mtp}}(\textrm{nbh},\overline{\theta})
\end{equation} for neighborhoods.

Thus, for each sample, a row vector of length $n$, $b(\textrm{X})$, can be formulated. The row vectors of $m$ samples are then compiled into a matrix $B$.

\[
B=
\begin{bmatrix} 
    b_{1}(\text{X}_1)  & b_{2}(\text{X}_1)  & \cdots & b_{n}(\text{X}_1) \\
    b_{1}(\text{X}_2)  & b_{2}(\text{X}_2)  & \cdots & b_{n}(\textrm{X}_2) \\
    \vdots & \vdots & \ddots & \vdots \\ 
    b_{1}(\text{X}_m)       & b_{2}(\text{X}_m)  & \dots & b_{n}(\text{X}_m) 
\end{bmatrix}
\]
When there are more training samples than there are parameters ($m>n$), $B$ becomes a tall matrix--the information matrix. The MTP maximizes the absolute value of the determinant of a $n\times n$ submatrix, $A$, comprising select rows of $B$. The $n$-member subset of samples forming $A$ is the active set. Conceptually, $A$'s absolute determinant represents the hypervolume of the $n$-parallelotope formed by the vectors of $A$'s rows. If the replacement of a member of the active set with a candidate sample would increase this hypervolume, then this candidate should be added to the training set.

Specifically, for candidate sample $X_\textrm{cand}$, we find $c_i(X_\textrm{cand})$ for all $n$ members of the active set, which represents the factor by which the absolute determinant would change should the $i$\textsuperscript{th} member of the active set by replaced with $X_\textrm{cand}$. A vector of these factors for each member in the active set is given by:

\begin{equation}
c(X_\textrm{cand})=(c_1(X_\textrm{cand}) \cdots c_n(X_\textrm{cand}))
\end{equation}

And is calculated by:
\begin{equation}
    c(X_\textrm{cand}) = \left( c_1(X_\textrm{cand}) \cdots c_n(X_\textrm{cand}) \right) = \left( b_1(X_\textrm{cand}) \cdots b_n(X_\textrm{cand}) \right) A^{-1}
\end{equation}

\begin{figure}[]
	\centering 
 
	\includegraphics[width=1\linewidth]{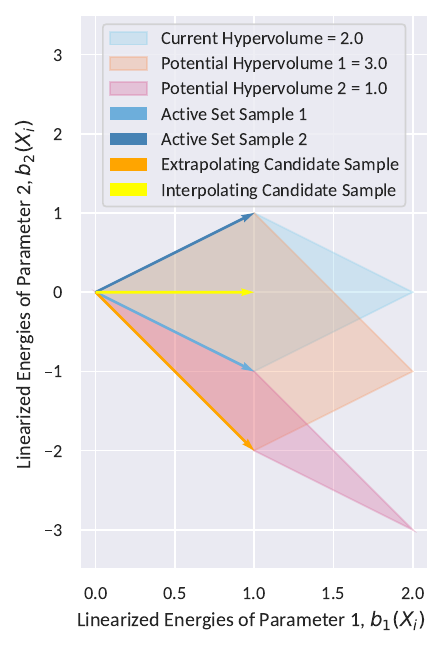}	
	\caption{This diagram illustrates MaxVol-based UQ for a case of two learnable parameters ($n=2$). The vectors of linearized energies of both samples in the active set (shades of blue) form the current hypervolume (blue area). The uncertainties of the two candidates are being evaluated. Should the orange sample replace Active Set Sample 1, the new hypervolume would become the orange area; should the orange sample replace Active Set Sample 2, the new hypervolume would become the pink area. Since the largest of these possible hypervolumes (orange) is larger than the current hypervolume (blue), the orange sample is considered to be extrapolating. The ratio of the largest possible hypervolume (orange) with the current hypervolume is the extrapolation grade, $\gamma(X_\textrm{orange}) = 3/2$. The yellow sample is interpolating, $\gamma(X_\textrm{yellow})=0.5$, and its possible hypervolumes aren't shown. In a typical query strategy, candidate samples are added to the training set when their extrapolation grade surpasses a user-defined threshold. } 
	\label{fig:example}%
\end{figure}

The extrapolation grade is defined as the maximum change in absolute determinant (hypervolume) that occurs when any single member of the active set is replaced by a candidate: $\gamma(X_\textrm{cand}) = \| c(X_\textrm{cand}) \|_{\infty}$. A visual example for $n=2$ is shown in Figure \ref{fig:example}.

If $\gamma(X_\textrm{cand})<1$, the configuration is considered an interpolation. If $\gamma(X_\textrm{cand})>1$, the configuration is considered an extrapolation. In MLIP-3 on-the-fly MD simulations, two control hyperparameters $\gamma_\textrm{select}$ and $\gamma_\textrm{break}$ thresholds are available to the user. Depending on the UQ mode, each time step yields either one configuration candidate or multiple neighborhood candidates. Candidates with extrapolation grades higher than $\gamma_\text{select}$ are selected, evaluated with \textit{ab initio} code, and added to the training set, replacing the corresponding member of the active set. Former active set members are kept in the training set.  When $\gamma_\text{break}$ is exceeded, the current MD simulation is terminated immediately, to be restarted after the MTP is retrained on the latest training set. 

Overall, one obtains Algorithm \ref{alg:al} which is roughly the methodology prescribed by the MLIP packages \cite{novikov2020mlip, podryabinkin2023mlip}. Parallelization across several different MD simulation conditions can accelerate the construction of the training set by boosting the sampling rate of candidates.

\begin{algorithm}
\caption{Bootstrapped On-The-Fly Learning with MD}
\label{alg:al}
\begin{algorithmic}[1]
\Require Curated initial training set $T$, MD simulation of interest $M$, untrained MTP of desired level $\text{mtp}$

\Loop
    \State $\text{mtp} \gets \text{mtp}$, fitted to $T$
    \State Initialize $S \gets \emptyset$
    
    \State Run MD simulation $M$ with $\text{mtp}$ from $t = t_0$
    \If{$M$ completes} 
        \State \textbf{break} 
    \EndIf
    \State $S \gets$ selected candidates of simulation $M$
    
    \State $D \gets$ Non-redundant selections of $S$
    \ForAll{selection $d \in D$}
        \State $d' \gets d$, evaluated with \textit{ab initio}
        \State Update training set $T \gets T \cup \{d'\}$
    \EndFor
\EndLoop
\end{algorithmic}
\end{algorithm}

Often, only a few atoms within a configuration are surrounded by a new environment. In neighborhood mode, MLIP-3 can extract these environments and pass the resulting cluster of atoms to the \textit{ab initio} code. Alternatively, neighborhood mode can pass the full configuration containing the selected neighborhoods to the \textit{ab initio} code and add it to the training set. This second approach is similar to configuration mode, except that instead of considering the extrapolation grade of the whole configuration, all local neighborhoods within that configuration are individually considered, which increases sensitivity.

In this study, we do not extract clusters since they are larger and thus more expensive than the periodic small cells. Neighborhood mode is solely used to increase sensitivity to new environments. 

\subsection{Small-Cell Training}
Here we outline our small-cell MTP training protocol for our two case studies. In the potassium system, we use a simple and deterministic protocol to better characterize performance and costs across a range of hyperparameters. In the sodium-potassium system, we generalize our approach and include randomized MD parameters, improving sampling and handling multiple elements. The computational details then follow.

\subsubsection{For a Simple System}
Much like earlier active learning small-cell studies, our basic small-cell active learning protocol proceeds in sequential stages with increasing simulation cell size at each stage. However, during each stage, multiple parallel active learning MD instances across a range of conditions are conducted. The active learning MD instances return selected configurations ($\gamma>\gamma_\textrm{select}$), which are pooled. When all instances terminate, redundant configurations are removed using MLIP-3, and the remainder are evaluated with DFT code and added to the training set. The potential is then refitted. This constitutes a single active learning iteration.

Next, more active learning MD instances are spawned, and the process is repeated until all of the MD instances can reach completion without selecting a configuration ($\forall\gamma,\gamma<\gamma_\textrm{select}$). Note that this is different from Algorithm~\ref{alg:al} where training is terminated after completing MD without breaking ($\forall\gamma,\gamma<\gamma_\textrm{break}$). At this point, the protocol proceeds to the next stage, where the size of the simulation box is increased and more active learning iterations are performed.

In this case study, each active learning MD instance is initially a perfect BCC crystal which is equilibrated in NVT for 100 ps, with a thermostat. The cell size progression starts with a single 2-atom BCC unit cell, which is replicated as outlined in Table~\ref{tab:stages}. A set of 24 parallel MD instances is used with varying combinations of temperatures and strains. Temperatures of 100, 300, 400, and 800~K were considered, as well as 6 uniformly spaced hydrostatic strains in the range of $\pm$5\%. For reference, potassium has a $T_\textrm{melt}$ of 336.5~K~\cite{lide2004crc}. Prior to active learning, we prepared an initial training set consisting of 20 primitive cells of $\pm$25\% uniformly spaced hydrostatic strains. 

\begin{figure*}[]
	\centering 
	\includegraphics[width=1\linewidth]{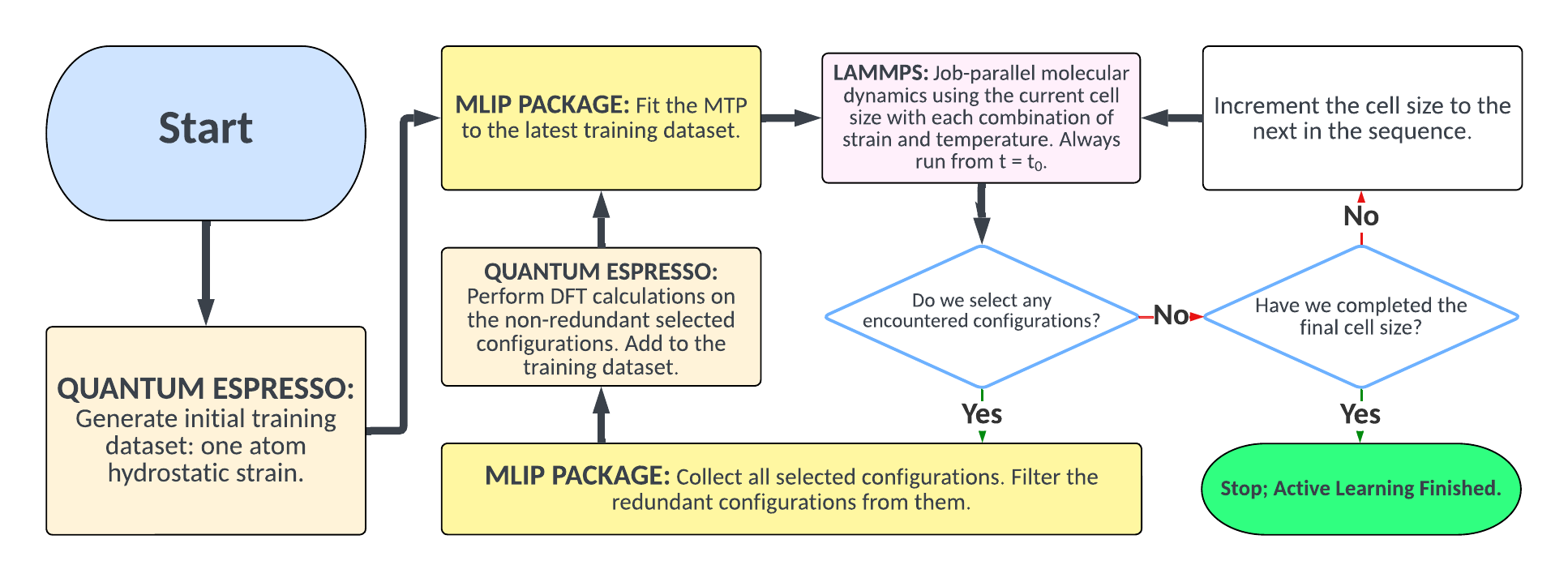}	
	\caption{A flowchart of the small-cell active learning protocol is shown. The colors and bold lettering indicate the corresponding software packages.} 
	\label{fig:flowchart}%
\end{figure*}

The simple training protocol is outlined in Figure~\ref{fig:flowchart}. We generated MTP potentials with MTP levels 8, 10, 12, 14, 16, and 18, as defined in Ref.~\cite{novikov2020mlip,podryabinkin2023mlip}. These MTP levels determine the expressiveness and the number of parameters of the MTP, relating to the basis function parameter count exponentially.

Some of the potassium potentials also involved sparsification of the training set. In these cases, the potential is refitted only to the active set. Removed configurations primarily consisted of smaller cells. Sparsification reduces the density of training configurations within the convex hull and increases the relative weight of presumably large-cell configurations. 

For our potassium potentials, in addition to benchmarking all UQ methods available in MLIP-3 and 6 different MTP levels, variations of the small-cell protocol described above were considered, such as the cell sizes used. Since many combinations are possible, we use a systematic notation as follows: 

{ \centering
  \textbf{aaabb(cc-dd)S} 
  \par}
with:
\begin{description}
    \itemsep-0.25em 
    \item \textbf{aaa}: Uncertainty quantification mode: CFG (configuration mode) or NBH (neighborhood mode)
    \item \textbf{bb}: MTP level
    \item \textbf{cc}: Lower bound atom count
    \item \textbf{dd}: Upper bound atom count
    \item \textbf{S}: Whether the training set is sparsified to only the active set after active learning.
\end{description}

For example, CFG08~(02-54) denotes a potassium MTP of level 8 with configuration mode small-cell active learning using stages 1-6 of Table~\ref{tab:stages}, inclusive. Notably, the upper atom count refers to the maximum active learning simulation size performed. Active learning MD does not necessarily select configurations at all cell sizes, especially larger ones.

\subsubsection{Generalizing for a More Complex System}
We then generalize our small-cell active learning protocol to more complex systems, such as multi-component alloys. The core procedure remains the same—iterative stages of parallel MD instances, configuration selection, and retraining—but key modifications improve robustness and the diversity of samples.

We first replace the previous grid of temperatures and initial strains with randomized values, sampled uniformly within the same ranges. We extend this framework to multi-component systems by incorporating variable concentrations and introducing an isotropic barostat (0-5000 MPa) to accommodate volume fluctuations across diverse compositions, both parameters being randomly sampled. High-temperature trajectories tend to yield more selections later into the protocol. Thus, we set around 25\% of the 24 active learning MD instances to the maximum temperature in the specified range. For any given stage, different cells of similar DFT cost—for instance, FCC, BCC, or even structures like surfaces—could be sampled from too, although we did not study this option. Overall, this generalized small-cell protocol adapts aspects of the earlier random sampling small-cell methods with active learning small-cell approaches.

We first apply this improved protocol to sodium and potassium individually, with the same cell size progression as in Table~\ref{tab:stages}. We then combine them, ensuring strong monoatomic performance and allowing each monoatomic MTP to train in parallel. When combining, we merge the training sets to form a new initial training set before resuming the protocol, with the same cell size progression in Table~\ref{tab:stages}, using perfect BCC crystals with random species. The species probabilities are chosen to target random concentrations but are restricted to ensure no active learning runs are monoatomic. A single NaK potential of level 18 is prepared with configuration mode UQ.

Notably, sodium-potassium is known to form a Na$_2$K binary phase (Pearson hP12, Strukturbericht C14). While not recommended in practice, this training protocol is designed agnostic of such a phase.

\subsection{Computational Details}
Convergence analysis of the chosen DFT parameters is available in~\ref{sec:app2}. For potassium, the chosen kinetic energy cutoff for wavefunctions was 60 Ry, and the kinetic energy cutoff for charge density and potentials was 300 Ry; for sodium, it was 90 Ry and 450 Ry. A 0.01 Ry Gaussian smearing was applied. The generalized gradient approximation functional by Perdew-Burke-Ernzerhof~\cite{perdew1996generalized} was employed, using pseudopotentials from Standard Solid-State Pseudopotentials and PSlibrary (K.pbe-spn-kjpaw\_psl.1.0.0.UPF, Na.pbe-spn-kjpaw\_psl.1.0.0.UPF)~\cite{prandini2018precision, dal2014pseudopotentials} was chosen. A Monkhorst-Pack uniform grid of k-points was used as detailed in Table \ref{tab:stages}.

\begin{table}[]
  \centering
  \caption{The size of the cells used in active learning MD simulations at each stage, measured in repetitions of the 2-atom BCC unit cells in each dimension. The k-points for each case are also listed.}
  \begin{tabular}{c|ccccc} 
  \hline 
  Stage & $X$ & $Y$ & $Z$ & Atom Count & k-points\\ 
  \hline 
  1 & 1 & 1 & 1 & 2 & $8 \times 8 \times 8$\\
  2 & 1 & 1 & 2 & 4 & $8 \times 8 \times 4$\\
  3 & 1 & 1 & 3 & 6 & $8 \times 8 \times 3$\\
  4 & 1 & 2 & 2 & 8 & $8 \times 4 \times 4$\\
  5 & 2 & 2 & 2 & 16 & $4 \times 4 \times 4$\\
  6 & 3 & 3 & 3 & 54 & $3 \times 3 \times 3$\\
  \hline
  \end{tabular}
  \label{tab:stages}
\end{table}

We employed the MTP as implemented in MLIP-3~\cite{podryabinkin2023mlip}. The chosen radial basis set comprised Chebyshev polynomials (RBChebyshev) with the default radial basis set sizes. The default fitting weights are used. The extrapolation grade thresholds were set to the recommendation of $\gamma_\textrm{select}=2.1$ and $\gamma_\textrm{break}=10$. Both configuration and neighborhood UQ modes were considered. The lower radial cutoff was chosen to match the minimum pair distance present in the training set. The upper cutoff was set to 7~\AA, based on convergence testing. Model weights at the start of the protocol were initialized deterministically with MLIP-3.

Fitting an MTP is a non-linear optimization process; the optimized MTP and its physical behavior depend on which local minimum the optimizer converged to. The active learning process yields a complete training set and one MTP fitted through the protocol. For benchmarks needing LAMMPS simulations, this fitted MTP is used. For the other benchmarks needing static predictions, we prepare MTP ensembles to address the sensitivity of the results to the choice of initial MTP parameters before optimization. Each ensemble comprises 24 MLIPs with random initial parameters fitted to the same complete training set. 

For all NVT MD simulations, both for active learning and benchmarking, we use a Langevin thermostat ~\cite{schneider1978molecular} with a dampening of 100 fs. All NPT simulations use a Nosé–Hoover thermostat and barostat \cite{hoover1985canonical}; damping parameters were 100 fs for temperature, and 1 ps for pressure. A 1 fs timestep is always used, and active learning MD instances all ran for 100 ps.

We used Quantum Espresso~6.6~\cite{giannozzi2009quantum,giannozzi2017advanced} for DFT calculations, LAMMPS~(23 Jun 2022)~\cite{thompson2022lammps} for MD simulations, and MLIP-3~(07 Jun 2023)~\cite{podryabinkin2023mlip} for MTP training; OpenMPI (4.0.3) and the GCC compiler (9.3.0) are used. Python (3.8.2), and various packages~\cite{harris2020array,Hunter:2007} were employed. Our Python script implementation and sample potentials are available on Github~\footnote{\href{https://github.com/RichardZJM/Small-Cell-MTP-Training}{Small-Cell MTP Training GitHub Repository}}. The scripts were designed for our case studies and may require modification to work well on alternative software and systems. The computer cluster ran Rocky Linux 8.10 (Green Obsidian) on Intel$^\text{®}$ Xeon$^\text{®}$ E5-2650 v4 CPUs in dual socket nodes with 256 GB of memory. Memory usage was substantially more than the available cache in all DFT and MD calculations. Since MLIP-3's fitting and filtering don't report CPU time, wall times were multiplied by core count usage to estimate CPU time. See \ref{sec:implementation} for more details.

\section{Results}

\subsection{Potassium Computational Cost}
In this subsection, the compute costs of the simple potassium small-cell protocol are reported. We consider both the wall time: the elapsed time measured for a floating allocation of cores; and the CPU time: the total time across all CPU cores spent actively executing the protocol. Both metrics exclude overheads such as queue times or file management. Notably, all DFT and MD costs are reported on 1 core—wall times can thus be improved with data parallelism, and CPU times have no parallelization inefficiency. The costs we present are:

\begin{enumerate}
    \item The protocol's distribution of costs across different processes (MD, DFT, fitting, filtering).
    \item The total DFT costs.
    \item The distribution of costs, training configurations, and active learning iterations across the stages of the protocol.
    \item The speedups of the small-cell protocol versus a large-cell approach using 54 atoms.
    \item DFT Task Parallelism.
\end{enumerate}

\begin{figure}[]
	\centering 
	\includegraphics[width=\linewidth]{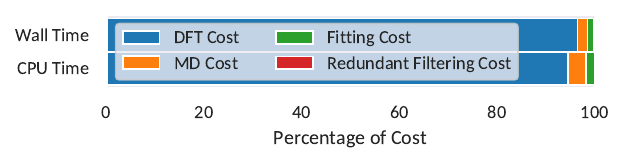}	
        \includegraphics[width=\linewidth]{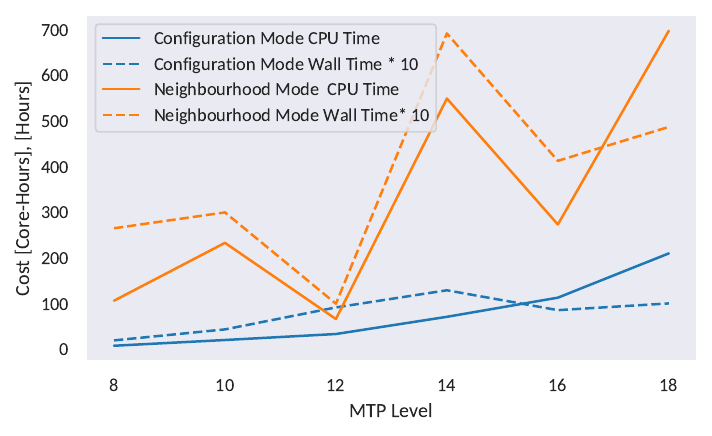}	
        \includegraphics[width=\linewidth]{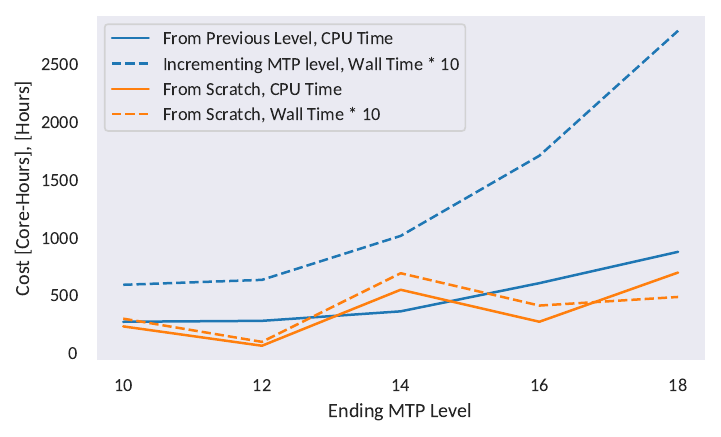}	
	\caption{(Top) The distribution of computational cost for training MTPs using small-cell training in neighborhood mode is averaged over MTP levels 8, 10, 12, 14, 16, and 18 and is plotted. The MD and DFT tasks used one core, MLIP-3 MTP fitting used 12 cores, and redundant configuration filtering used 2 cores.
 \\ (Middle) The DFT costs of our small-cell protocols with all stages up to 54 atoms are shown. We assess  MTP levels 8, 10, 12, 14, 16, and 18, with both configuration and neighborhood modes. MLIP-3's neighborhood mode yielded larger and more unpredictable computational costs, selecting an unpredictable amount of 16- and 54-atom configurations for training.
 \\ (Bottom) A comparison of the CPU and wall costs of training neighborhood MTPs using our small-cell protocol. From scratch refers to training to a given MTP level using a user-curated training initial training set. From previous level refers to the cumulative cost of training from MTP 8, and repeatedly incrementing the MTP level by 2 and applying the small-cell protocol on the training set of the preceding level.
 } 
	\label{fig:distribution}%
	\label{fig:cpuCost}%
        \label{fig:merge}
\end{figure}

The computational cost distribution of small-cell training is shown in Figure \ref{fig:distribution}. The figure was prepared by averaging the training cost distribution of NBH08(2-54), NBH10(2-54), NBH12(2-54), NBH14(2-54), NBH16(2-54) and NBH18(2-54).  DFT calculations account for about 95~\% of the overall training cost, both in wall time and CPU time. The active learning MD runs and fitting costs account for the balance, while filtering costs are negligible. Although not shown for brevity, configuration mode has a similar cost distribution.

We then investigate the DFT costs of MTPs who trained with all atom sizes (2-54). The results are shown in Figure~\ref{fig:cpuCost}. There is a general increase in CPU cost with MTP level. Neighborhood mode's costs are larger as it consistently selects 16- and 54-atom cells. In configuration mode, 16-atom cells were selected only once; 54-atom cells were never selected. These large cells contribute disproportionately to the overall cost.

In MLIP training, it may be necessary to expand the parameter count of an existing potential by retraining with a larger model. For active learning, this usually requires expanding the training set since a new model size may yield new uncertainties. Figure~\ref{fig:merge} shows the cumulative cost of training an MTP to a given end level by incrementally running small-cell active learning using the final training set of a previous level, starting from MTP 8. Overall, there are few cost benefits to such an approach in small-cell training. Although it may reduce the overall number of training configurations, it often induces a greater number of larger cells, which disproportionately increases the computational cost.  

\begin{figure*}[]
	\centering 
	\includegraphics[width=\textwidth ]{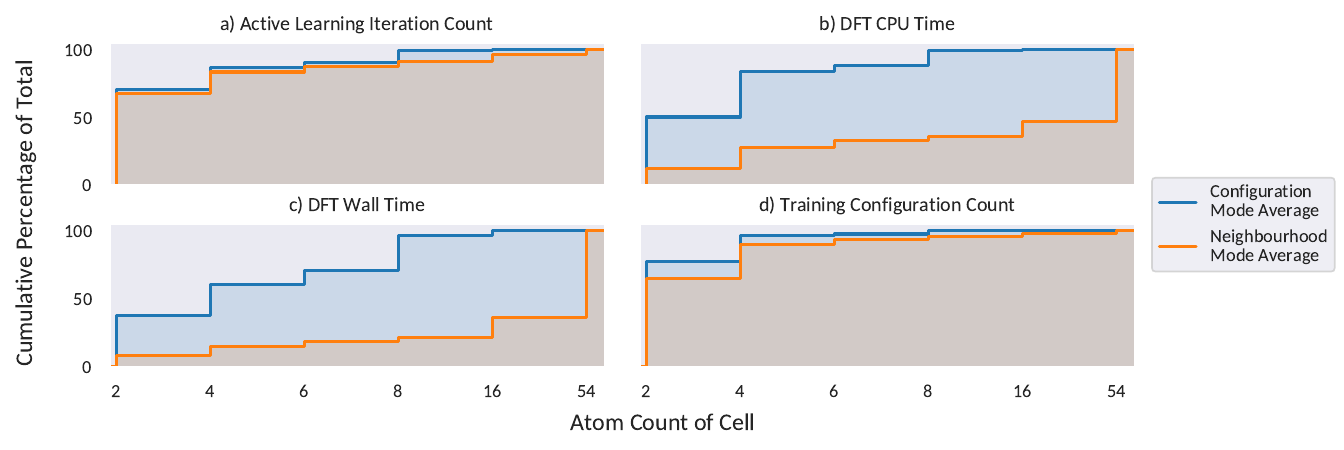}	
	\caption{A cumulative percentage plot describing a) Active Learning Iteration Count, b) DFT CPU Time Cost, c) DFT Wall Time Cost, d) Training Configuration Count. Results are shown by stage (atom count) and are averaged over small-cell protocols using MTP levels 8, 10, 12, 14, 16, and 18 for neighborhood and configuration mode. } 
	\label{fig:stages}%
\end{figure*}

Next, we break down the costs and training set composition by UQ mode in Figure~\ref{fig:stages}. Most selections (\ref{fig:stages}d) and active learning iterations (Figure~\ref{fig:stages}a) involve small cells. However, the few larger cells included in the training set led to disproportionately large costs (Figure~\ref{fig:stages}b,c). We observe a higher proportion of smaller cells in configuration mode (Figure~\ref{fig:stages}d) and note no significant differences with the MTP level. 

\begin{figure}[]
	\centering 
	\includegraphics[width=\linewidth]{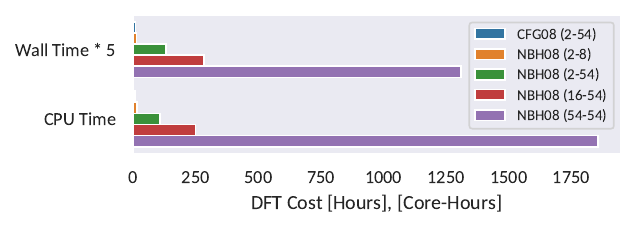}
        \includegraphics[width=\linewidth]{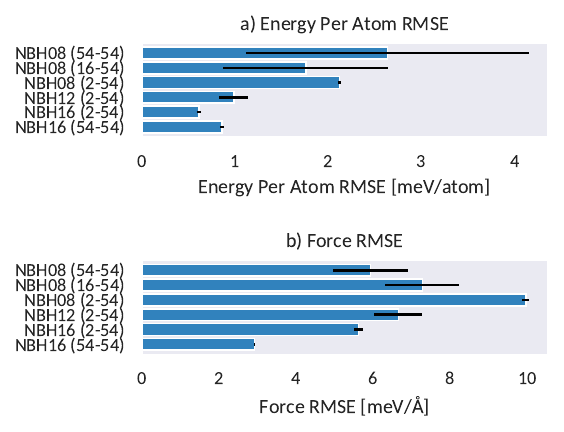}
	\caption{(Top) DFT costs of small-cell training and large-cell training are compared for a level 8 MTP. Large-cell training (NBH08 (16-54), NBH08 (54-54) costs more than small-cell training (CFG08 (2-54), NBH08 (2-8), NBH08 (2-54)). For example, NBH08 (54-54) costs 237$\times$ more CPU time than CFG08 (2-54) 
 \\(Bottom) The training RSME of neighborhood mode small-cell ensembles (NBH08 (2-54), NBH12 (2-54), NBH16 (2-54)) is compared with large-cell ensembles (NBH08 (16-54), NBH08 (54-54), NBH16 (54-54)). Higher MTP levels lead to smaller force and energy training errors. Small-cell training leads to smaller energy training errors and larger force training errors. 1 standard deviation is shown.
 }
	\label{fig:schemes}%
        \label{fig:trainingRSME}
\end{figure}

We then compare our small-cell method against approaches with 16- then 54-atom cells (NBH08(16-54)), and exclusively 54-atom cells (NBH08(54-54)) in Figure~\ref{fig:schemes}. Our corresponding small-cell method (NBH08(2-54)) yields a 9.9$\times$ wall, 17.4$\times$ CPU speedup versus 54-atom cells (NBH08(54-54)). These speedups are similar to Meziere~\textit{et al.}'s 17.5$\times$ CPU time speedup using small cells versus a 48- and 54-atom cell approach ~\cite{meziere2023accelerating,liu2021alpha}. CFG08(2-54) offers a 133.8$\times$ wall, 237$\times$ CPU cost speed-up as compared to NBH08(54-54). Speedups for MTP level 12 and 16 are also available in Table~\ref{tab:costs}.
 
 Cells including 8 atoms are notable as the largest cell typically selected in configuration mode, making CFG08(2-54) functionally identical to CFG08(2-8). 8-atom cells were also used in Meziere~\textit{et al.} \cite{meziere2023accelerating}, Pickard's boron potential~\cite{pickard2022ephemeral} and the GTTP \cite{PhysRevB.107.125160}.  When we limit neighborhood mode's upper bound of the cell sizes to 8 atoms (NBH08(2-8)), we observe DFT speedups of 95.5$\times$ wall, 119.4$\times$ CPU. 

\begin{table}[]
    \centering
    \caption{DFT speedups of small-cell active learning are compared to active learning on only 54-atom cells. 8- and 54-atom upper-bound cell sizes for MTP levels 8, 12, and 16 in neighborhood mode are shown. Small-cell MTP level 12 exhibits higher speedups than other MTP levels since it did not involve selecting any 54-atom cells.}
    \begin{tabular}{cc|ccc}
     \cline{3-5}
     & & \multicolumn{3}{c}{MTP Level} \\ \hline
    \multicolumn{2}{c|}{Cell Sizes} & 8 & 12 & 16 \\ \hline
    \multirow{2}{*}{CPU Time} & $\leq$ 8 Atoms (2-8) & 119.4 & 141.5 & 112.0 \\ 
    & $\leq$ 54 Atoms (2-54) & 17.4 & 99.4 & 62.0 \\ \hline
    \multirow{2}{*}{Wall Time} & $\leq$ 8 Atoms (2-8)  & 95.5 & 113.5 & 73.3 \\ 
    & $\leq$ 54 Atoms (2-54) & 9.9 & 55.3 & 17.0 \\ \hline
    \end{tabular}
    \label{tab:costs}
\end{table}

\begin{figure}[]
	\centering 
	\includegraphics[width=\linewidth]{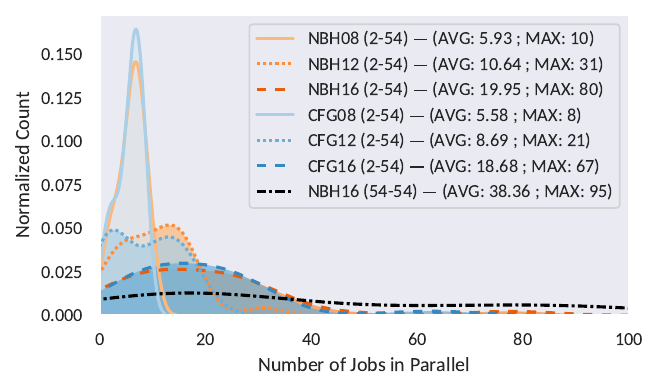}	
	\caption{Kernel density estimate of the number of DFT jobs (selections) in active learning iterations. Average and maximum are shown.} 
	\label{fig:coreCount}%
\end{figure}

The number of selections in any given iteration of active learning is not constant, making active learning a variable core workload that may be difficult to implement efficiently in a typical fixed-resource allocation. To help further characterize these variable core workloads, we provide a kernel density estimate of selections of some training sessions in Figure~\ref{fig:coreCount}. There tends to be a larger number of selected configurations in a few active learning iterations near the end of each stage, yielding larger maximum counts. 

\subsection{Potassium Force and Energy Predictions}
For brevity, we focus accuracy comparisons on MTPs trained using neighborhood mode; trends for configuration mode are similar. We consider the accuracy of the following metrics:

\begin{enumerate}
    \item Training errors for both forces and energies.
    \item Test force residuals.
    \item The effects of sparsification on test force residuals.
    \item The effect of incrementally introducing larger cells on test force.
\end{enumerate}

We first consider the training root square mean errors (RSME) of small-cell ensembles (NBH08(2-54), NBH12(2-54), NBH16(2-54)) and larger-cell ensembles (NBH08(16-54), NBH08(54-54), NBH16(54-54)) in Figure~\ref{fig:trainingRSME}. Larger MTP levels lead to smaller force and energy training errors. Small-cell training leads to smaller energy training errors and larger force training errors than large-cell training. 

Using \textit{ab initio} MD, we generate 20 liquid DFT configurations of 128 atoms at 600~K with no strain as a test set. Notably, 128-atom cells are larger than all the configurations in all training sets.

\begin{figure}[]
	\centering 
	\includegraphics[width=\linewidth]{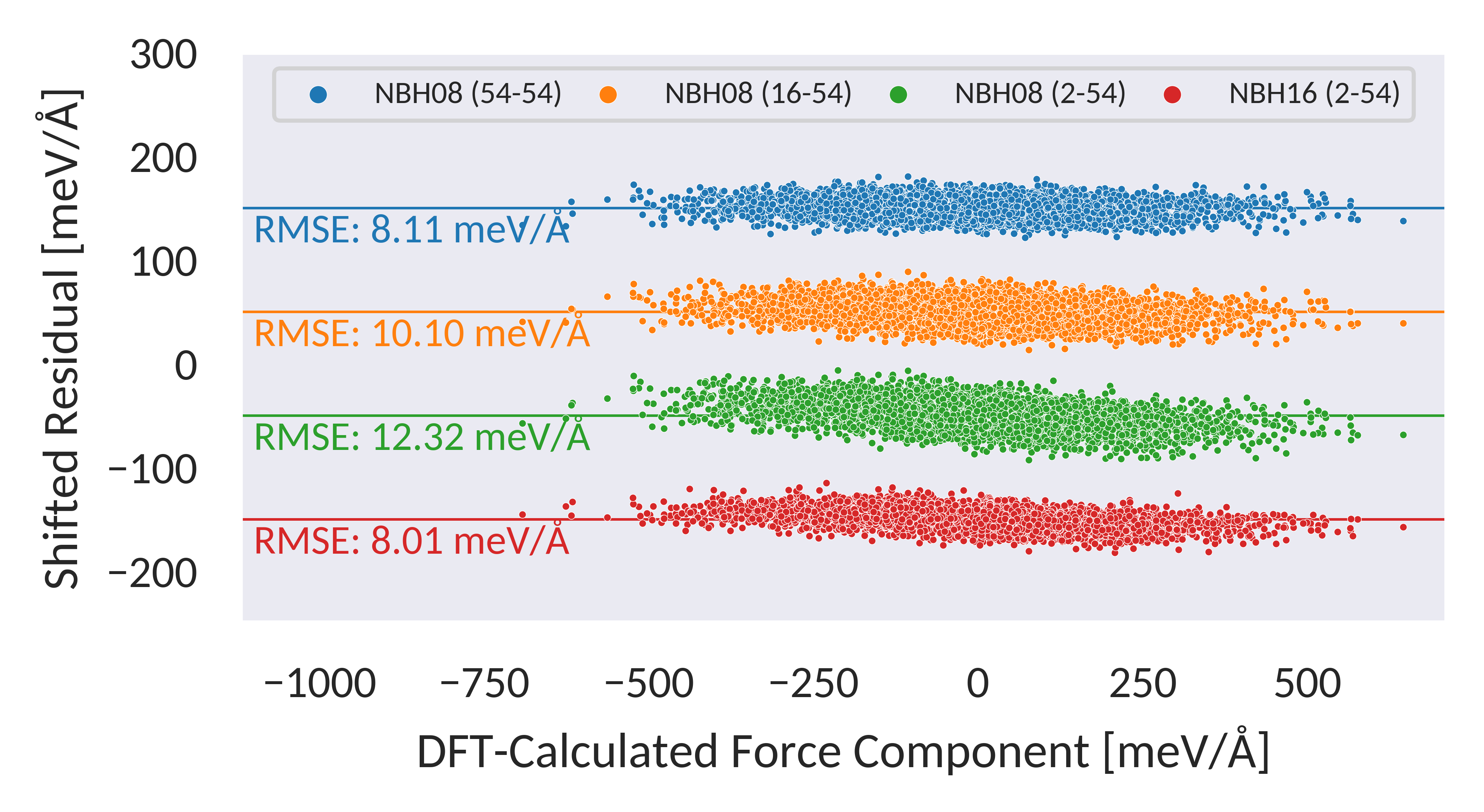}	
	\caption{The force component residuals of neighborhood mode small-cell-trained MTP ensembles (NBH08(2-54)); (NBH16(2-54)) are compared with larger-cell-trained MTP ensembles (NBH08(16-54), NBH08(54-54)). Small cells bias towards zero force and increase test force RSME.} 
	\label{fig:testForce}%
\end{figure}

In Figure~\ref{fig:testForce}, the force residuals of small-cell-trained MTP ensembles (NBH08(2-54)), (NBH16(2-54)) are compared with larger-cell-trained MTP ensembles (NBH08(16-54), NBH08(54-54)). Higher levels and larger cells yield lower force RSMEs. A slight negative slope in the residuals of small-cell ensembles indicates a minor bias towards zero force, which is alleviated in higher MTP levels. Test energy RSME is similar in both small- and large-cell approaches.

\begin{figure}[]
	\centering 
	\includegraphics[width=\linewidth]{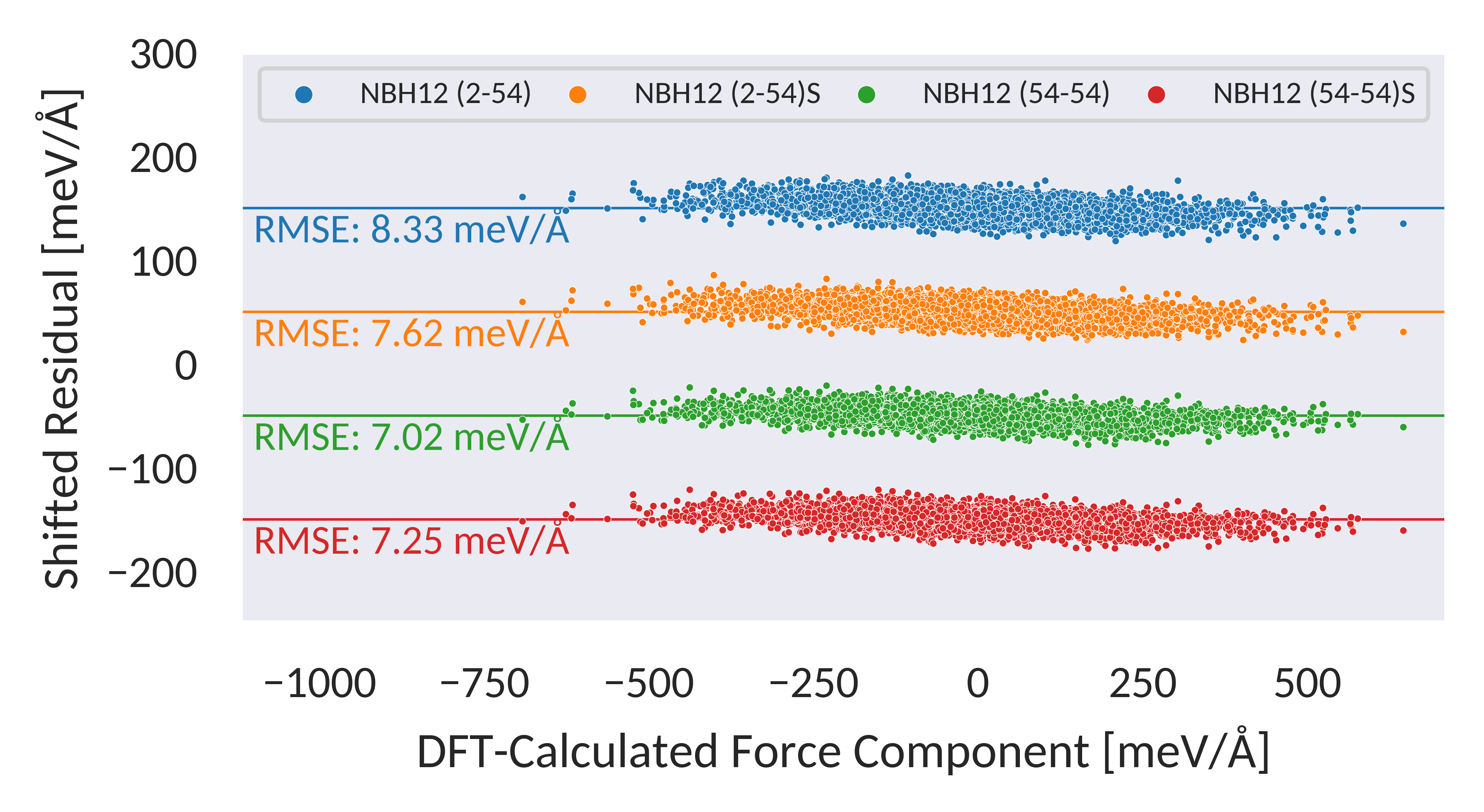}	
	\caption{The effects of sparsification on the force component residuals is shown for small-cell ensembles (NBH12(2-54), NBH12(2-54)S) and large-cell ensembles (NBH12(54-54), NBH12(54-54)S). Sparsification improves small-cell ensembles and worsens in large-cell ensembles.} 
	\label{fig:sparseForce}%
\end{figure}

Noting these small-cell biases, we consider the effects of training set sparsification, in which we remove all training set members outside the active set. These removed configurations are mostly smaller cells. In Figure~\ref{fig:sparseForce}, residuals are shown for small-cell ensembles (NBH12(2-54), NBH12(2-54)S) and large-cell ensembles (NBH12(54-54), NBH12(54-54)S). We observe force RSME and bias improvements from using sparsification on small-cell ensembles and greater errors when large cells are used. Sparsification is less beneficial with higher MTP levels.

\begin{figure}[]
	\centering 
	\includegraphics[width=\linewidth]{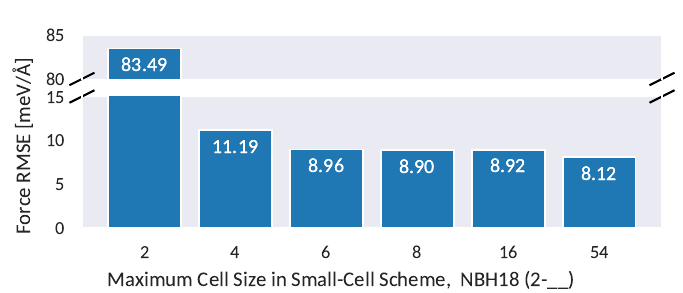}	
	\caption{ Diminishing returns in test force component RSME of the small-cell protocol with increasing upper bound atom count. }
	\label{fig:progressiveStage}%
\end{figure}

Using the same test set, we consider the effects of incrementally including larger cells into a small-cell protocol in Figure~\ref{fig:progressiveStage}; a neighborhood mode MTP of level 18 is considered.

\subsection{Solid Potassium Properties}
We evaluate several solid properties using MTPs and compare them with DFT values:

\begin{enumerate}
    \item BCC Equation of state.
    \item Lattice parameters: BCC, FCC, SC, and HCP.
    \item Energies: BCC, FCC, SC, and HCP.
    \item Elastic constants: $c_{11},c_{12},c_{44}$
    \item Unstable stacking fault energy (USFE) in the full BCC slip system ($\left(110\right)\left[\overline{1}11\right]$).
    \item Vacancy formation energy.
    \item Vacancy migration energy.
\end{enumerate}

\begin{figure}[]
	\centering 
	\includegraphics[width=\linewidth]{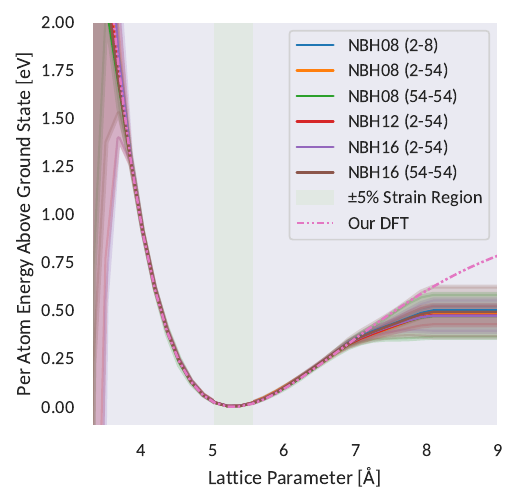}	
	\caption{ Equation of state (BCC) as predicted by our MTP ensembles and DFT. The $\pm$5\% hydrostatic strain range used in active learning simulations is highlighted. The MTP upper cutoff is 7~\AA. 2 standard deviations shaded. }
	\label{fig:eos}%
\end{figure}

We begin with the BCC equation of state (EOS) in Figure~\ref{fig:eos}. From 4 to 7~\AA, the DFT and MTP agreement is excellent. The upper cutoff of the MTPs was 7~\AA, at which point the MTP prediction diverges from the DFT benchmark.

\begin{figure*}[h!]
	\centering 
	\includegraphics[width=\textwidth ]{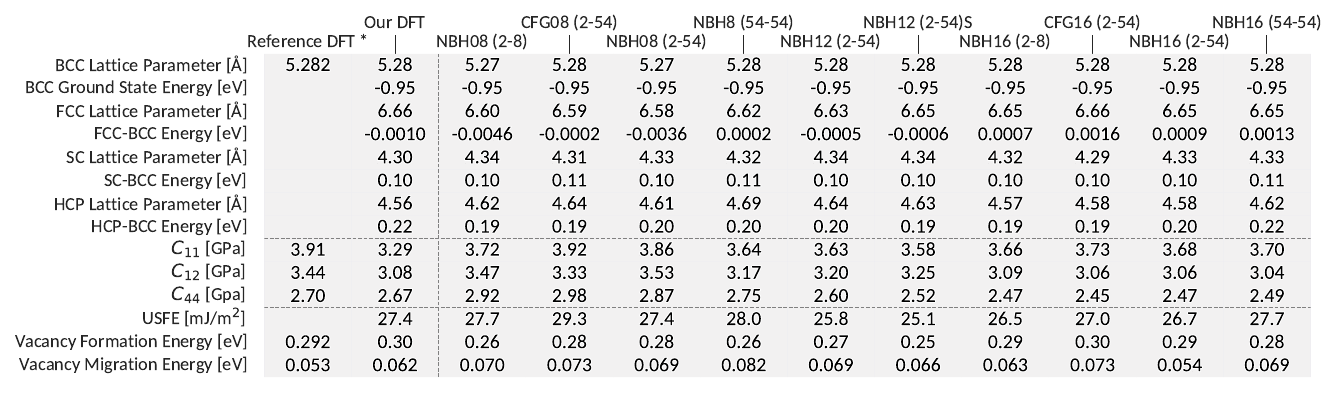}	
	\caption{Predicted properties of solid potassium. Ensembles of 24 MTPs were used for all but the last three predictions, which used LAMMPS simulations. Uncertainties are not shown to avoid clutter—they are of the order of $10^{-4}$~\AA\, for lattice parameters, $10^{-5}$~eV for lattice energies, and 0.1~GPa for elastic constants. * Ma~{\it et al.}~\cite{ma2019effect}.} 
	\label{fig:heatmap}%
\end{figure*}

We now briefly explain how we calculate the remaining solid properties. The same method is used for DFT and MTP, although the next benchmarks use an MTP ensemble. The ground state lattice parameters and energy for BCC, FCC, SC, and HCP are calculated by fitting energies to a 3\textsuperscript{rd}-order Birch-Murnaghan equation of state~\cite{birch1947finite}. Energies are then expressed relative to BCC. We then calculate 0~K elastic constants with strains up to $\pm$1\%, adapting the energy-strain method from LePage and Saxe~\cite{le2001symmetry}. We emphasize that these DFT elastic reference values are calculated with the same parameters as those used during active learning. They are converged relative to per-atom energies, but not converged relative to elastic constants.

The remaining solid properties use a single MLIP potential, rather than the 24-MLIP ensembles. We first assess the USFE in the full BCC slip system ($\left(110\right)\left[\overline{1}11\right]$) by adapting the z-relax slab model~\cite{sholl2022density}. 10 layers were used, and the outermost two layers on each side of the slab were fixed. We also benchmark the vacancy formation energy using strain-free energy minimization on a 53-atom BCC cell. Vacancy migration is compared, as calculated per the nudged elastic band method on relaxed neighboring vacancies~\cite{ma2019effect,henkelman2000climbing}. 

The results are presented in Table~\ref{fig:heatmap}. MTP predictions are in good agreement with DFT results and are fairly consistent across MTPs. However, the scatter and errors of the elastic constants are larger than for other properties, and MTP values deviate from the DFT values.

Elastic constants depend on higher-order energy derivatives, which make them more sensitive to small differences in the potential energy surface. Accordingly, additional DFT calculations of elastic constants were carried out, with higher accuracy converged for elastic constants (\ref{sec:app2}). A comparison is presented in Table~\ref{tab:elastic}. Notably, all of the MTPs of level 16 or higher agree with the higher-accuracy elastic-converged DFT, even though they were trained using lower-accuracy, energy-converged DFT. Training cell size has no or minimal effect. 

To ensure convergence during MTP training, another CFG16 (2-54) was trained with higher DFT accuracy, doubling the energy cutoffs and k-points. This resulted in no notable changes in elastic predictions or other properties. 

\begin{table}[]
\centering
\caption{The effects of DFT accuracy on elastic predictions [GPa] are tabulated for DFT and for MTPs trained with the specified DFT accuracy throughout active learning. Ordered by accuracy, the DFT parameters used are: Energy-Converged (EC)$\rightarrow$ Increased Accuracy (IA) $\rightarrow$ Elastic-Converged (CC).}
\begin{tabular}{r|ccc}
\hline
 & $C_{11}$  & $C_{12}$  & $C_{44}$  \\ \hline
  Reference DFT (Ma et al. \cite{ma2019effect}) & 3.91 & 3.44 & 2.70 \\ \hline
  This Work, DFT (EC) & 3.29 & 3.08 & 2.67 \\ 
  This Work, DFT (CC) & 3.73 & 3.06 & 2.45 \\ \hline
  NBH16 (2-8) (EC) & 3.66 & 3.09 & 2.45 \\ 
  CFG16 (2-54) (EC) & 3.73 & 3.06 & 2.45 \\ 
  NBH16 (2-54) (EC) & 3.68 & 3.06 & 2.47 \\ 
  NBH16 (54-54) (EC) & 3.70 & 3.04 & 2.49 \\ \hline
  CFG16 (2-54) (IA) & 3.70 & 3.07 & 2.46 \\ \hline
\end{tabular}
\label{tab:elastic}
\end{table}

\subsection{Liquid Potassium Properties}
\begin{figure}[]
    \centering
    \includegraphics[width=\linewidth]{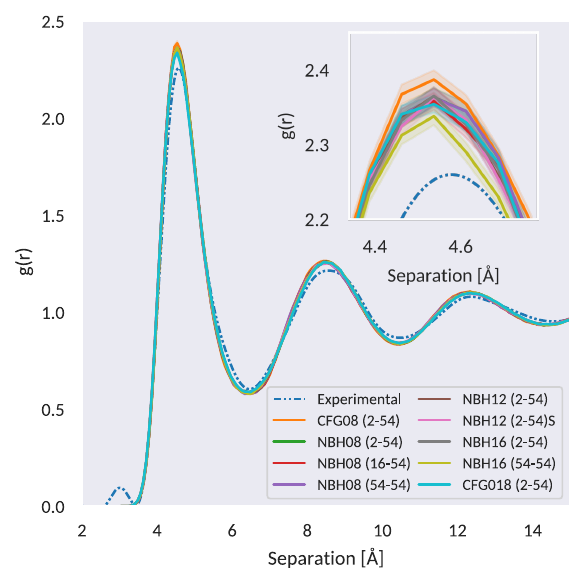}
    \includegraphics[width=\linewidth]{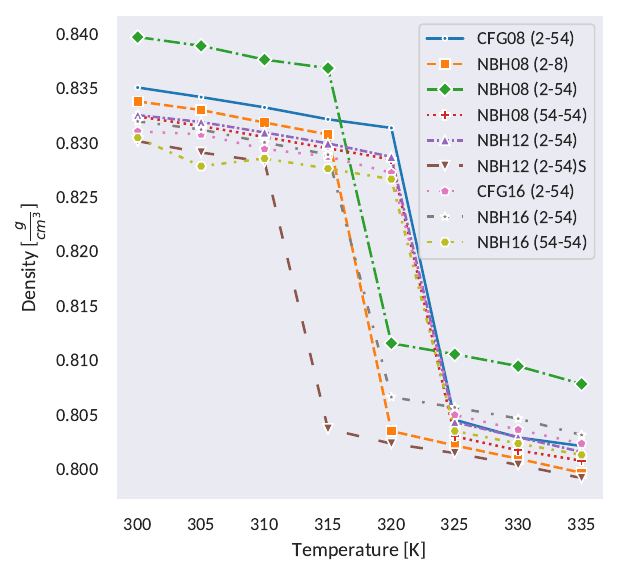}
    
    \caption{ (Top) Potassium radial distribution function predictions at 408~K, 5.40~\AA\, are shown and compared with X-Ray diffraction by Greenfield \cite{greenfield1971x}. A minor artifact at low g(r) remains from converting Greenfield's structure factors to RDFs. The inset magnifies the first RDF peak. 2 standard deviations of bootstrapped error are shaded. 
     \\(Bottom) A density-temperature plot is used to calculate the melting point with the NPT-interface method \cite{zhang2012comparison}. The experimental melting point is 336.65 K \cite{lide2004crc}. Miryashkin et al. using MTPs obtained $309.52\pm0.92$~K \cite{miryashkin2023bayesian}.}
    \label{fig:melt}
    \label{fig:rdf}
\end{figure}

We first present the radial distribution function (RDF) of liquid potassium at 408~K and 5.40~\AA\ as predicted by the MTPs in Figure~\ref{fig:rdf}. All models agree well with each other and fairly well with X-ray diffraction results from Greenfield ~\cite{greenfield1971x}.

Next, we calculate melting points using the NPT-interface method~\cite{zhang2012comparison} with the resultant density-temperature plot in Figure~\ref{fig:melt}. Our predictions vary between 315 and 325~K, while the experimental melting point is 336.65~K~\cite{lide2004crc}. Miryashkin {\it et al}. report $309.52\pm0.92$~K using MTP active learning~\cite{miryashkin2023bayesian}.

During these two-phase simulations, which used 2048 atoms, the maximum observed extrapolation grade was $\gamma_{max}\approx2.56$ for MTPs trained using active learning up to 54 atoms, slightly above the selection threshold of $\gamma_\text{select}=2.1$. When using configuration mode, no encountered configurations were selected: $\forall\gamma, \gamma<\gamma_\text{select}=2.1$.

We then computed the critical exponents of specific heat at constant volume, $C_v$, in the ordered phase $\alpha'$, for CFG08 (2-54) and NBH08 (54-54), for various system sizes, $L$, at a near-melting density of $\rho = 0.8325$ g cm$^{-3}$. For each MTP, temperature, and system size, we equilibrated a cubic cell of side length $L$, for 4 ns, and computed the $C_v$ over 1 ns using the energy fluctuations method. A power-law fit was performed on the reduced temperatures based on the observed critical temperature, $T_c$, yielding the critical exponents in Table \ref{tab:critexp}. 

Near $T_c$, systems exhibit large-scale fluctuations. Critical exponents are highly sensitive to these effects, and the agreement between models suggests similar predictions for these long-range fluctuations. For $L=20$, we also observe within-error agreement with the theoretical value for the universality class of three-dimensional Ising-like systems: $\alpha = 0.110 \pm 0.003$ \cite{sengers1986thermodynamic, pelissetto2002critical}. The critical exponents are expected to be the same for the ordered phase ($\alpha'$) as the unordered phase ($\alpha$) for this system, $\alpha$ = $\alpha'$.

\begin{table}[]
\centering
\caption{The critical exponents of the specific heat, $C_v$, in the ordered phase, $\alpha'$, of a small- and large-cell trained MTP, calculated for different cubic cells of side length $L$ at $\rho = 0.8325$ g cm$^{-3}$. Standard error and $R^2$ are shown. The theoretical value for the critical exponent is $0.110 \pm 0.003$ \cite{sengers1986thermodynamic} }
\begin{tabular}{c|cccc}
\hline
  MTP & $L$ [Cells] & $T_c$ [K] & $\alpha'$ &$R^2$\\
\hline
\multirow{3}{*}{\shortstack{NBH08\\(54-54)}} & 10 & 372.5 & 0.124 $\pm$ 0.006 & 0.918 \\
 & 15 & 371.0 & 0.125 $\pm$ 0.005 & 0.927 \\
 & 20 & 369.5 & 0.107 $\pm$ 0.005 & 0.918 \\
\hline
\multirow{3}{*}{\shortstack{CFG08\\(2-54)}} & 10 & 377.0 & 0.127 $\pm$ 0.004 & 0.935 \\
 & 15 & 376.0 & 0.118 $\pm$ 0.005 & 0.915 \\
 & 20 & 373.5 & 0.112 $\pm$ 0.005 & 0.916 \\
\hline
\end{tabular}
\label{tab:critexp}
\end{table}

\subsection{Sodium-Potassium}
We begin with a cost breakdown of our more complex small-cell protocol for sodium-potassium in Table \ref{tab:nakcosts}. K and Na were first trained separately, and their combined training set became the initial training set for NaK. The costs are broken down by step accordingly; configuration mode UQ and MTP level 18 are used. 

Notably, DFT makes up only 48\% CPU and 63.5\% wall of the final training costs. This lower percentage is mostly caused by increased fitting costs, which resulted from slower convergence, especially in NaK. Indeed, most fitting iterations reached MLIP-3's default cap of 1000 optimizer iterations, which suggests that new selections heavily alter the objective space; more cores (23 instead of 12) were also used for fitting, potentially affecting parallel efficiency. Increased MD costs are also observed since the barostat and random concentrations induce selections later into active learning MD runs. 12 cores were used for filtering instead of 2. Nonetheless, the total costs are still lower than the DFT costs of a level 8 potassium MTP trained on large cells (NBH08 (54-54)). Only 15 16-atom configurations were sampled; no 54-atom configurations were sampled.

\begin{table}[]
\centering
\caption{Small-cell training cost breakdown for training K and Na individually, followed by combining them into NaK. * The overall wall time if Na and K were trained in parallel.}
\begin{tabular}{c|c|cc}
\hline
\multirow{2}{*}{Model} & \multirow{2}{*}{Category} & \multicolumn{2}{c}{Cost [core-hrs][hrs]} \\
\cline{3-4}
 & & CPU & Wall \\
\hline
\multirow{5}{*}{Potassium} & MD & 64.4 & 3.1 \\
 & DFT & 253.2 & 15.5 \\
 & Filtering & 0.5 & 0.04 \\
 & Fitting & 30.5 & 1.3 \\ \cline{2-4}
 & Total & 348.5 & 20.0 \\
\hline
\multirow{5}{*}{Sodium} & MD & 107.0 & 6.2 \\
 & DFT & 141.7 & 13.2 \\
 & Filtering & 0.3 & 0.03 \\
 & Fitting & 63.0 & 2.7 \\ \cline{2-4}
 & Total & 312.0 & 22.2 \\
\hline
\multirow{5}{*}{Sodium-Potassium} & MD & 127.2 & 8.2 \\
 & DFT & 414.8 & 44.5 \\
 & Filtering & 0.5 & 0.04 \\
 & Fitting & 470.4 & 20.5 \\ \cline{2-4}
 & Total & 1012.8 & 73.1 \\
\hline
\multirow{5}{*}{Total} & MD & 298.6 & 17.4 \\
 & \textbf{DFT} & \textbf{809.7} & \textbf{73.2} \\
 & Filtering & 1.2 & 0.1 \\
 & Fitting & 563.8 & 24.5 \\ \cline{2-4}
 & \textbf{Total} & \textbf{1673.4 } & \textbf{95.2* / 115.2} \\
\hline
\end{tabular}
\label{tab:nakcosts}
\end{table}

We then validate our potential on a selection of benchmarks: 
\begin{enumerate}
    \item Na$_2$K C14 Laves phase equation of state.
    \item Liquid eutectic densities at selected temperatures, 1 bar.
    \item Liquid eutectic specific heat, $C_p$, at 20 $^{\circ}$C, 1 bar.
    \item Average equimolar radial distribution function.
    \item Melting point at eutectic composition.
\end{enumerate}
All eutectic benchmarks were performed at the experimental concentration of 67.3$\,$at.\% potassium (Na$_{0.327}$K$_{0.673}$)\cite{leonchuk2022nak}.

\begin{figure}[]
	\centering 
	\includegraphics[width=\linewidth]{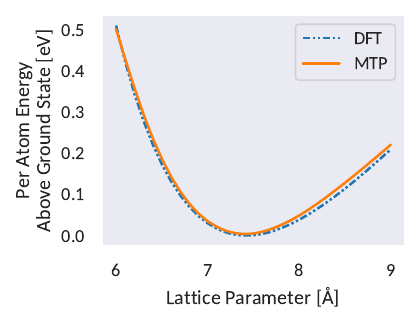}	
	\caption{ The Na$_2$K C14 Laves phase equation of state, MTP and DFT. }
	\label{fig:nakeos}%
\end{figure}

We begin with the C14 equation of state in Figure \ref{fig:nakeos}, which is a fair match, even though the potential was trained agnostic of the C14 phase. Fitting to a 3\textsuperscript{rd}-order Birch-Murnaghan equation of state~\cite{birch1947finite}, we obtain a lattice constant of 7.411~\AA\ at -1.140 eV/atom versus 7.424 \AA\ at -1.143 eV/atom per DFT. Fit errors are negligible. 

In a 200 K, 1 bar MD simulation of a 1536-atom cell of perfect C14 Na$_2$K, we observed a maximum selection grade of $\gamma_{max}\approx2.7$, slightly above the selection threshold of $\gamma_\text{select}=2.1$.

Next, we benchmarked against the room temperature, equimolar, average RDF from Henninger et al.'s neutron diffraction data \cite{henninger1966atomic}. Their reported room temperature was taken to be 22.5 $^{\circ}$C and 1 bar was assumed; the average RDF was calculated from the weighted average of partial RDFs. The results are shown in Figure \ref{fig:nakeos}.

\begin{figure}[]
    \centering
    \includegraphics[width=\linewidth]{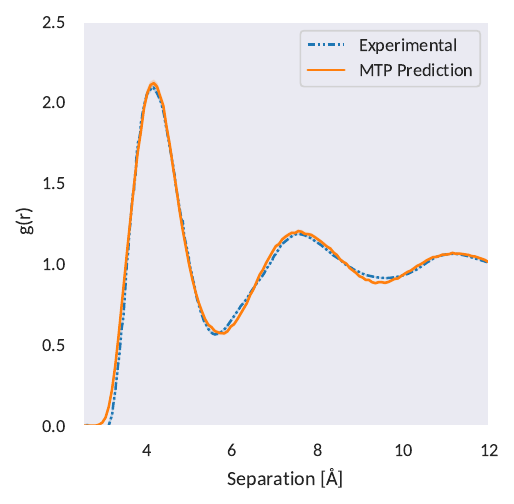}
    \caption{Average radial distribution function predictions of equimolar NaK at 22.5 $^{\circ}$C, 1 bar, are compared with neutron diffraction data at room temperature by Henninger et al. \cite{henninger1966atomic}. 2 standard deviations of bootstrapped error are shaded for both. Experimental errors only include uncertainty from density calculations.}
    \label{fig:nakrdf}
\end{figure}

We then calculate the 1-bar density of liquid potassium at 20, 100, and 550 $^{\circ}$C, comparing them against experimental values from Leonchuk et al. \cite{leonchuk2022nak} in Table \ref{tab:dens}. Notably, 550 $^{\circ}$C is outside the training temperature range used. We also compare our results for the specific heat at constant pressure for 1 bar and 20 $^{\circ}$C against experimental values from Foust \cite{foust1976sodium}. Adapting the finite difference method, we perform a linear regression on enthalpies obtained in intervals of 2 from 282 to 304 K. We predict a $C_p$ of 954 $\pm$ 11\ versus an experimental value of 971 J kg$^{-1}$ K$^{-1}$.

\begin{table}[]
\centering
\caption{Experimental (Leonchuk et al. \cite{leonchuk2022nak}) and predicted densities, with 95\% confidence interval, of eutectic NaK in g/cm$^3$. }
\begin{tabular}{c|cc}
\hline
Temperature [$^{\circ}$C] & Experimental & Predicted \\
\hline
20  & 0.867 & 0.852  $\pm$ 0.004 \\
100 & 0.855 & 0.833  $\pm$ 0.005 \\
550 & 0.749 & 0.737  $\pm$ 0.008 \\
\hline
\end{tabular}
\label{tab:dens}
\end{table}

\begin{figure}[]
    \centering
    \includegraphics[width=\linewidth]{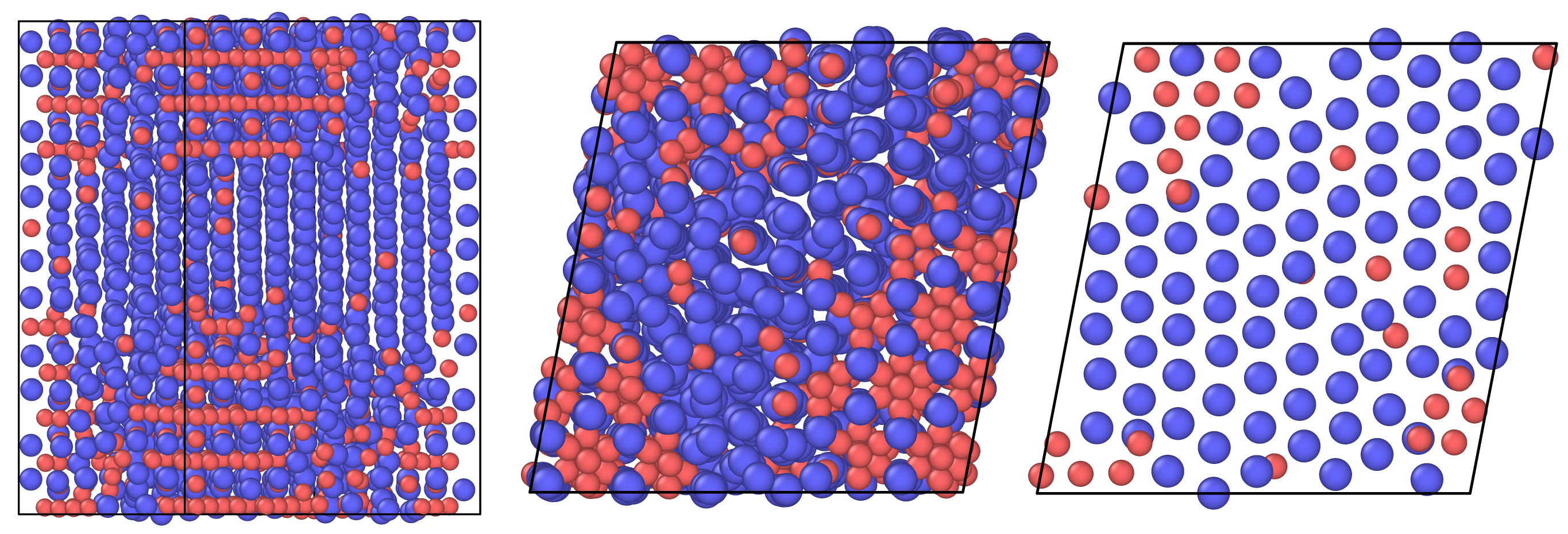}
    \caption{Predicted solid eutectic NaK structure (Na$_2$K (C14) + K (BCC)). Oblique side view drawn at half scale (left); a view normal to the interface (middle); a 3 $\AA$ thick slice normal to the interface, taken at the center of the cell, highlighting BCC (right). Na is red; K is blue; atoms drawn to 65\% of experimental atomic radius for clarity.}
    \label{fig:nakc14}
\end{figure}

Finally, we calculate the melting point at the eutectic concentration using the NPT-interface method. To prepare the solid half of the interface, we begin with a C14 Laves structure and randomly assign atomic species such that an eutectic concentration is reached. We then use a hybrid Metropolis Monte Carlo (MC) \cite{sadigh2012scalable} and molecular dynamics approach in NPT at 150 K with 10 type swaps per MD step and an MC scaling temperature of 240 K, until converged. The resultant solid structure is rendered with OVITO \cite{stukowski2009visualization}, with the solid half of the (0001) interface shown in Figure \ref{fig:nakc14}. A C14 phase Na$_2$K and a BCC potassium phase are clearly visible, and are confirmed through OVITO's common neighbor analysis. 

Monitoring density-temperature, we observe the melting point between 245 and 250 K, which is in good agreement with experimental value of 260.5 K \cite{leonchuk2022nak}, and fair agreement with an earlier \textit{ab initio} NaK phase diagram study by Huan \textit{et al.} \cite{huang2022ab}, which predicts 268 K. We note that in a sample interface simulation at 290K, no encountered configurations were selected: $\forall\gamma, \gamma<\gamma_\text{select}=2.1$. Costs were too high to enable active learning for all melting simulations.

Notably, the interfaces do not recrystallize. Instead, the density discontinuity is where the interface stops moving completely. In smaller-scale tests, the potential also has an issue crystallizing, instead favoring an amorphous phase which visually resembles that observed by Reitz~\textit{et al.} \cite{reitz2019simulating}, \cite{reitz2017monte}. Their melting point prediction of 250-260 K also agrees with our findings. We confirmed that both the MTP and DFT (up to 88 atoms) energetically favor the crystalline structure over the amorphous phase, with strong agreement.

We also augmented the training protocol with active learning on C14 cells up to 48 atoms; both Na$_2$K and random concentrations are tested. Recrystallization issues persist with no substantial change to melting point predictions.

\section{Discussion}
\label{sec:disc}
These case studies demonstrate how small-cell training can lead to substantial computational savings as compared to large-cell training. This is despite fairly simple training protocols, which were not tailored for the phenomena being benchmarked. One can train a potential adequate for many simulations in liquid and solid potassium in just a day on consumer computer hardware, for sodium-potassium in a week. Further wall time speedups are available with data parallelism of DFT jobs—here, only 1 core was used to facilitate cost comparison. Small cells also use much less memory, lowering the barrier to entry and enabling more hardware configurations. Small-cell training stands to yield even more drastic speedups with more expensive and accurate methodologies such as coupled cluster and quantum Monte Carlo calculations, which tend to scale more poorly than DFT with atom count. Moreover, even our larger training cells are of modest size compared to our cutoff radius, which can include over 50 atoms. For comparison, Jinnouchi~\textit{et al} used up to 144-atom interfaces~\cite{jinnouchi2019fly} and the extracted clusters in the MLIP-3 paper reached 168 atoms with a 5~\AA\ upper MTP cutoff~\cite{podryabinkin2023mlip} (compared to the 7~\AA\ cutoff used here).

As expected, training and validation errors improve with increasing MTP levels. Curiously, small-cell trained potentials often have smaller training energy errors than their larger-cell counterparts, as shown in Figure~\ref{fig:trainingRSME}. The opposite is observed for force errors. In the case of the MTP and most other MLIPs, energy is often calculated on a per-atom basis and summed. Since DFT does not provide per-atom energies, there arises a source attribution problem where there is ambiguity as to how individual atom energies contribute to the whole. Since small cells have fewer atoms, the source attribution problem is mitigated. This effect may explain why larger cells exhibit high training energy error with higher error variance.

Small-cell training not only leads to worse test force RSME but also introduces a bias towards zero force, as can be seen in Figure~\ref{fig:testForce}. This bias may be caused by an excess of crystal-like structures formed by periodicity in small cells, exhibiting lower forces. One mitigating strategy is an increase in MTP level, although partial sparsification of the training set or augmentation with a small number of larger cells can both be considered.

The MTPs were unable to accurately describe potassium's cohesive energy (see Figure \ref{fig:eos}), a limitation of training with solely bulk materials and a finite, 7~\AA\, upper cutoff. This issue would also affect surface energies and could be addressed by including surfaces in the training protocol. Another possibility is to first fit a pair potential able to handle longer-range interactions cheaply \cite{nitol2023hybrid}, such as empirical or tabulated potentials. The MLIP can be overlaid to these pair potentials to capture mid-range, many-body interactions that are missed by the pair potential. For short-range interactions, the Ziegler-Biersack-Littmark (ZBL) potential \cite{ziegler1985stopping} can be used as a baseline \cite{byggmastar2019machine,wang2019deep,wood2017quantum}. While implementing such a hybrid potential was out of scope for this study, we recommend its consideration in future work for both short-range and long-range interactions.

Remarkably, the MTPs led to elastic constants with higher accuracy than the DFT upon which it was trained. The errors of the energy-converged DFT are likely averaged over the entire training set—taken at many different temperatures and stress states—onto the MTP basis functions, which can achieve a good physical description of elastic behavior in potassium.

The recrystallization issue with the NaK potential is likely an erroneous energy barrier caused by insufficiently diverse sampling of phenomena during training, and not strictly training cell size, although the former often requires the latter. Manually curated interfaces or a more discriminatory UQ metric might be needed since new selections were not observed during the sample interface simulations. Another possibility is that it may take longer to form C14, which was not explored in this study. Although unlikely, since MTP predictions agree with DFT in the amorphous and crystalline phases, we do not rule out DFT as the root cause.

More generally, small-cell training is constrained by the UQ methodology. UQ does not necessarily predict error. In most schemes, it estimates the variance of the MLIP for a candidate configuration, which often but not necessarily correlates to the error. Most UQ methods depend on the MLIP descriptor and thus act like a projection of higher-dimensional configurational space to a lower-dimensional one. The UQ methodology may indicate that two configurations are nearby in the lower-dimension latent embedding space even if they would have been far away in the higher-dimension configurational space. Despite these concerns, small-cell MTPs were successful with all UQ modes available in the MLIP-3 package. Other UQ methods likely exhibit performance characteristics, especially since MaxVol measures uncertainty based on an outer hyperboundary. Refs. \cite{annevelink2023statistical} and \cite{kulichenko2024data} provide a more complete look at other UQ metics.

Similar considerations arise when generalizing to other MLIP formalisms, even though small-cell approaches have seen previous success with other potentials \cite{pickard2022ephemeral,PhysRevB.107.125160}. In theory, our small-cell protocol is generalizable to local MLIPs with robust UQ in periodic boundaries. In practice, neural network potentials tend to require significantly more samples than the linear regression of the MTP basis set and may expose insufficient exploration in more naive protocols. Highly non-linear regressors need to be well tested before using MaxVol UQ. Message-passing potentials such as MACE \cite{batatia2022mace} are also ill-suited for a small-cell approach, since they lack the same notion of a cutoff radius. The MLIP formalisms immediately amenable to small-cell training would be SNAP \cite{thompson2015spectral} and ACE \cite{drautz2019atomic}, the latter of which has MaxVol UQ implemented in LAMMPS.

To generalize the small-cell training protocol proposed here to other materials and contexts, consider specializing the parallel, random MD trajectories. Our NaK training scheme gave reasonable C14 predictions without deliberately training for it. This may not be so in other systems. It is preferable to include structures known to be relevant in future work using small cells. For NaCl, Sun~\textit{et al.}~\cite{sun2024interatomic} included shear structures, chlorine dimers, and supersaturated solutions; Luo~\textit{et al.}~\cite{luo2023set} included different Zr crystal structures, point defects, surfaces, and extended defects. Random structure generators like USPEX \cite{lyakhov2013new} and RandSpg \cite{avery2017randspg} could help introduce more diverse MD conditions too. Moreover, protocols that explore the free energy space, such as parallel tempering~\cite{earl2005parallel, swendsen1986replica}, could allow autonomous exploration and characterization of thermodynamically important phases. Augmentation with a small number of larger cells can improve finite-size biases. Then, partial sparsification of the smaller cells in the training set can improve the relative weight of these larger cells.

In our study, configuration mode led to considerable savings versus neighborhood mode $\approx3\times$ CPU, $\approx5\times$ wall with minor accuracy changes. The larger cells that the neighborhood mode selects yield little test error improvement (see Figure~\ref{fig:progressiveStage}). If additional accuracy is needed after small-cell active learning, we could then consider on-the-fly learning on large-cell simulations of the phenomena of interest. In this context, small-cell methods can be considered for pretraining with later fine-tuning in these large simulations.

Importantly, these speedups are for training costs only. The computational cost of using the MLIP to run MD is independent of the training set and training methodology. As such, larger cells and neighborhood mode could be preferred when training costs represent a small proportion of an MLIP's lifespan. Conversely, small cells can pretrain potentials efficiently, allowing users to rapidly get good quality MLIPs ready for production runs, with the possibility of later augmentation.

We did not test the impact of changing the uncertainty threshold on MLIP training cost. It could be profitable to do so, possibly by using an adaptive uncertainty threshold that varies as the training advances.

Finally, we discuss the accuracy of small-cell training. A common doubt that can be raised about training with small cells is that many relevant phenomena and structures that are observed when simulating larger systems are absent if only a small number of atoms are considered. Examples include defect diffusion, plasticity, interfaces, and phase transitions. However, this doubt implies that MLIPs directly learn phenomena in large cells. This is incorrect. Rather, MLIPs learn about local environments and interactions. On-the-fly approaches using large-cell simulations of the phenomenon of interest simply ensure that local atomic environments in the training set match those observed during the simulation. In contrast, small-cell training involves diverse atomic environments—our study and previous small-cell training research suggest these environments have a sufficient overlap with those encountered during the larger-scale simulations. It should be pointed out that large-scale properties are emergent. For instance, the Ising model, which can be parametrized based on very short-range calculations, can capture arbitrarily long-range emergent behavior. 

We further contextualize small-cell methods by pointing out that a sensible initial training set generally reduces the number of configurations that must be added during on-the-fly MLIP training. Small cells offer a rapid and cost-effective way to generate such an initial training set.

\section{Conclusions}
 We have shown that small-cell training offers a powerful approach for the development of compatible MLIPs. Our conclusions are:

\begin{enumerate}
    \item Small-cell training enables up to two orders of magnitude of training cost reductions. 
    \item A MTP appropriate for potassium simulations took as little as 120 core-hours, 1700 for NaK.
    \item Small-cell training leads to smaller training energy errors, but larger training force errors, compared to large-cell training.
    \item Small-cell training introduces force biases toward zero force.   
    \item Physical properties, liquid and solid, were well described for both potassium and NaK.
    \item Potassium can be well described using MLIPs of low complexity, as compared to other materials and alloys. More sophisticated training, such as that used for NaK, or larger cells, may be needed in other systems.
    \item For small-cell training, configuration mode—as implemented in the MLIP-3 software package—was preferable to neighborhood mode. 
    \end{enumerate}

Small-cell training requires careful training regimen selection, including cell sizes, UQ, and simulation conditions. As general guidance, we recommend first applying small-cell active learning like a rapid pretraining method, possibly followed by fine-tuning during large-cell simulations of phenomena of interest. Future research may consider more complex systems, and explore the effect of different UQ metrics, MLIP descriptors, and MLIP regressors.


\section*{CRediT statement}
\textbf{Zijian Meng:} Conceptualization, Methodology, Software, Validation, Visualization, Investigation, Data Curation, Writing - Original Draft, Writing - Review \& Editing \textbf{Hao Sun:} Conceptualization, Methodology, Software, Writing - Review \& Editing \textbf{Edmanuel Torres:} Methodology, Writing - Review \& Editing \textbf{Christopher Maxwell:} Methodology, Writing - Review \& Editing \textbf{Ryan Grant:} Methodology, Resources, Writing - Review \& Editing, Supervision, Funding acquisition \textbf{Laurent Karim Béland:} Conceptualization, Methodology, Resources, Writing - Review \& Editing, Supervision, Funding acquisition, Project administration


\section*{Acknowledgments}
An acknowledgment to Matthew Thoms for insightful discussions.

\section*{Funding}
Financial support for this work was provided through the Natural Sciences and Engineering Research Council of Canada (NSERC), Mitacs, and the University Network for Excellence in Nuclear Engineering (UNENE). We also thank the Digital Research Alliance of Canada—formerly known as Compute Canada—and the Centre for Advanced Computing at Queen's University for the generous allocation of computer resources. This work was partly funded by Atomic Energy of Canada Limited, under the auspices of the Federal Nuclear Science and Technology Program.

\clearpage
\appendix

\section{DFT Convergence}
\label{sec:app2}
\subsection{Active Learning DFT Convergence}
For the training sets of our active learning protocol, we converged our DFT parameters using the per-atom energy to approximately 1 meV. Plane-wave kinetic energy cutoff converged at 60 Ry for K, and 90 Ry for Na, as shown in Table \ref{tab:ecutK} and \ref{tab:ecutNa}. Tables \ref{tab:k2}, \ref{tab:k4}, \ref{tab:k6}, \ref{tab:k8}, \ref{tab:k16}, and \ref{tab:k54} show k-point convergence by increasing cell size with potassium atoms.

\begin{table}[h!]
\centering
\caption{Potassium plane wave cutoff energy convergence relative to per atom energy. 60 Ry was ultimately used, slightly more than Ma \textit{et al.} \cite{ma2019effect}, much more than the recommended minimum of 41 Ry \cite{prandini2018precision}.}
\begin{tabular}{c c c c c c c c}
\hline
Cutoff [Ry] & 45 & 55 & 60 & 65 \\ 
\hline
Energy [eV/atom]  & -0.9425 & -0.9464  & -0.9475 &  -0.9478  \\
\hline
\end{tabular}
\label{tab:ecutK}
\end{table}

\begin{table}[h!]
\centering
\caption{Sodium plane wave cutoff energy convergence relative to per atom energy. 90 Ry was ultimately used much more than the recommended minimum of 66 Ry \cite{dal2014pseudopotentials}.}
\begin{tabular}{c c c c c c c c}
\hline
Cutoff [Ry] & 70 & 80 & 90 & 100 \\ 
\hline
Energy [eV/atom]  & -1.1943 & -1.2181  & -1.2251 &  -1.2258  \\
\hline
\end{tabular}
\label{tab:ecutNa}
\end{table}

\begin{table}[h!]
\centering
\small \setlength{\tabcolsep}{5pt}
\caption{K-point convergence for 2-atom cell, $1 \times 1 \times 1$ unit cells. $8 \times 8 \times 8$ Monkhorst-Pack grids were ultimately used.}
\begin{tabular}{c c c c c c c c}
\hline
K-points & $7 \times 7 \times 7$ & $8 \times 8 \times 8$ & $9 \times 9 \times 9$ & $10 \times 10 \times 10$ \\ 
\hline
Energy [eV/atom] & -0.9452  & -0.9475 & -0.9487 & -0.9475 \\
\hline
\end{tabular}
\label{tab:k2}
\end{table}

\begin{table}[h!]
\centering
\small \setlength{\tabcolsep}{5pt}
\caption{K-point convergence for 4-atom cell, $1 \times 1 \times 2$ unit cells. $8 \times 8 \times 4$ Monkhorst-Pack grids were ultimately used.}
\begin{tabular}{c c c c c c c c}
\hline
K-points  & $7 \times 7 \times 4$ & $8 \times 8 \times 4$ & $9 \times 9 \times 5$ & $10 \times 10 \times 5$ \\ 
\hline
Energy [eV/atom] & -0.9472  & -0.9475 & -0.9479 & -0.9475 \\
\hline
\end{tabular}
\label{tab:k4}
\end{table}

\begin{table}[h!]
\centering
\small \setlength{\tabcolsep}{5pt}
\caption{K-point convergence for 6-atom cell, $1 \times 1 \times 3$ unit cells. $8 \times 8 \times 3$ Monkhorst-Pack grids were ultimately used.}
\begin{tabular}{c c c c c c c c}
\hline
K-points & $7 \times 7 \times 3$ & $8 \times 8 \times 3$ & $9 \times 9 \times 3$ & $10 \times 10 \times 4$ \\ 
\hline
Energy [eV/atom] & -0.9457  & -0.9479 & -0.9487 & -0.9475 \\
\hline
\end{tabular}
\label{tab:k6}
\end{table}

\begin{table}[h!]
\centering
\small \setlength{\tabcolsep}{5pt}
\caption{K-point convergence for 8-atom cell, $1 \times 2 \times 2$ unit cells. $8 \times 4 \times 4$ Monkhorst-Pack grids were ultimately used.}
\begin{tabular}{c c c c c c c c}
\hline
K-points  & $7 \times 4 \times 4$ & $8 \times 4 \times 4$ & $9 \times 9 \times 5$ & $10 \times 10 \times 5$ \\ 
\hline
Energy [eV/atom] & -0.9473  & -0.9475 & -0.9474 & -0.9475 \\
\hline
\end{tabular}
\label{tab:k8}
\end{table}

\begin{table}[h!]
\centering
\small \setlength{\tabcolsep}{5pt}
\caption{K-point convergence for 16-atom cell, $2 \times 2 \times 2$ unit cells. $4 \times 4 \times 4$ Monkhorst-Pack grids were ultimately used.}
\begin{tabular}{c c c c c c c c}
\hline
K-points & $3 \times 3 \times 3$ & $4 \times 4 \times 4$ & $5 \times 5 \times 5$ & $6 \times 6 \times 6$ \\ 
\hline
Energy [eV/atom] & -0.9481  & -0.9475 & -0.9475 & -0.9477 \\
\hline
\end{tabular}
\label{tab:k16}
\end{table}

\begin{table}[h!]
\centering
\small \setlength{\tabcolsep}{5pt}
\caption{K-point convergence for 54-atom cell, $3 \times 3 \times 3$ unit cells. $3 \times 3 \times 3$ Monkhorst-Pack grids were ultimately used.}
\begin{tabular}{c c c c c c c c}
\hline
K-points & $2 \times 2 \times 2$ & $3 \times 3 \times 3$ & $4 \times 4 \times 4$ & $5 \times 5 \times 5$ \\ 
\hline
Energy [eV/atom] & -0.9481& -0.9487  & -0.9477 & -0.9476  \\
\hline
\end{tabular}
\label{tab:k54}
\end{table}

\subsection{Convergence Relative to Elastic Constants}
 We ultimately used $30 \times 30 \times 30$ k-points (the same as Ma et al. \cite{ma2019effect}) and a 200 Ry plane wave cutoff energy to converge the DFT parameters of a 2-atom cell relative to the elastic constants. Convergence data is shown in Table \ref{table:elasticEcut} for the cutoff energy and Table \ref{table:elasticK} for the k-points.

\begin{table}[h!]
\centering
\caption{Plane wave cutoff energy convergence relative to elastic constants in a 2-atom cell ($\approx$ 5.28 \AA). 200 Ry was ultimately used.}
\begin{tabular}{c c c c}
\hline
Cutoff [Ry] & 180 & 200 & 220 \\ 
\hline
$C_{11}$ [GPa] & 3.73 ± 0.02 & 3.73 ± 0.02 &  3.73 ± 0.02 \\
$C_{12}$ [GPa] & 3.07 ± 0.08 & 3.06 ± 0.08 &  3.06 ± 0.08 \\
$C_{44}$ [GPa] & 2.45 ± $\leq$0.01  & 2.45 ± $\leq$0.01  & 2.45 ± $\leq$0.01  \\
\hline
\end{tabular}
\label{table:elasticEcut}
\end{table}

\begin{table}[h!]
\centering
\caption{K-point convergence relative to elastic constants in a 2-atom cell ($\approx$ 5.28 \AA). $30 \times 30 \times 30$ Monkhorst-Pack grid were ultimately used.}
\begin{tabular}{c c c c}
\hline
K-points & $28 \times 28 \times 28$ & $30 \times 30 \times 30$  & $32 \times 32 \times 32$ \\ 
\hline
$C_{11}$ [GPa] & 3.63 ± 0.02 & 3.73 ± 0.02 & 3.71 ± 0.03 \\
$C_{12}$ [GPa] & 3.12 ± 0.08 & 3.06 ± 0.08 & 3.08 ± 0.07 \\
$C_{44}$ [GPa] & 2.48 ± $\leq$0.01 & 2.45 ± $\leq$0.01 & 2.45 ± $\leq$0.01 \\ 
\hline
\end{tabular}
\label{table:elasticK}
\end{table}

\section{Software Implementation Details}
\label{sec:implementation}
Most of the scripting was performed in Python on a computing cluster and packaged into a pip-compatible package\footnote{\href{https://github.com/RichardZJM/Small-Cell-MTP-Training}{Small-Cell MTP Training GitHub Repository}}. Overall, we provide our scripts as a reference for users seeking to perform MTP small-cell learning with MLIP-3 and Quantum Espresso, but not necessarily as plug-and-play software. These scripts can be easily adapted to perform conventional active learning as well.

We prepare two primary versions of the protocol, one designed to run on a fixed allocation of cores, branch \codeword{main}, and the other which uses a floating resource allocation, branch \codeword{floatingAllocation}. The former uses a primitive semaphore to schedule DFT jobs should their required core usage exceed the quantity of cores available; memory usage is unmanaged. In the latter, MD and DFT tasks each run as separate SLURM jobs. This requires a tolerant cluster policy, low queue times, and minimal node launch failures, which can cause deadlock.

Several subfolders are included in the package: \codeword{templates}, which provides constant parameters to generate tasks throughout the package; \codeword{io}, to read and write files in the required formats; \codeword{activeLearningSections}, to handle file transfers and call other software; and \codeword{ensembles}, to generate and manage MTP ensembles. Except for the last folder, these components work together in \codeword{activeLearnPotential.py} to fully automate the small-cell training protocol, following the hyperparameters outlined in a configuration JSON file.

The MLIP-3 package includes converters between its internal format and that of the Vienna Ab Initio Simulation Package. Since Quantum Espresso is used in this study, conversion is handled in the \codeword{io} modules, passing Python dictionaries of relevant parameters. Three notable processing steps occur during conversion: first, DFT energy values are shifted so that zero per-atom energy corresponds to that of an atom in a vacuum; second, the MLIP-3 internal configuration format requires virial stresses to be multiplied by the cell volume; third, conversion between 0- and 1-indexing of types is performed as needed. Unit conversion to eV and $\AA$ is also applied.

Running the \codeword{runActiveLearningScheme} function of \codeword{activeLearnPotential.py} follows the process described in the Methodology section. The final output is a directory tree containing the trained potential, the training configurations in the MLIP-3 internal format, all DFT jobs, and a log file with metrics obtained throughout active learning. An ensemble can be generated from a complete training set and used for configuration predictions, but not MD simulations, with functions in the \codeword{ensembles} module.

More information is available in the Github's README.

\section*{Data Availability Statement}
The scripts used for training and selected models with their training data are available from Github\footnote{\href{https://github.com/RichardZJM/Small-Cell-MTP-Training}{Small-Cell MTP Training GitHub Repository}}. Additional data such as benchmarking scripts may be provided upon request.
\newpage

\bibliography{example.bib}

\begin{thebibliography}{10}
\expandafter\ifx\csname url\endcsname\relax
  \def\url#1{\texttt{#1}}\fi
\expandafter\ifx\csname urlprefix\endcsname\relax\def\urlprefix{URL }\fi
\expandafter\ifx\csname href\endcsname\relax
  \def\href#1#2{#2} \def\path#1{#1}\fi

\bibitem{deringer2019machine}
V.~L. Deringer, M.~A. Caro, G.~Cs{\'a}nyi, Machine learning interatomic
  potentials as emerging tools for materials science, Advanced Materials
  31~(46) (2019) 1902765.

\bibitem{mishin2021machine}
Y.~Mishin, Machine-learning interatomic potentials for materials science, Acta
  Materialia 214 (2021) 116980.

\bibitem{mueller2020machine}
T.~Mueller, A.~Hernandez, C.~Wang, Machine learning for interatomic potential
  models, The Journal of chemical physics 152~(5) (2020).

\bibitem{behler2007generalized}
J.~Behler, M.~Parrinello, Generalized neural-network representation of
  high-dimensional potential-energy surfaces, Physical review letters 98~(14)
  (2007) 146401.

\bibitem{bartok2013representing}
A.~P. Bart{\'o}k, R.~Kondor, G.~Cs{\'a}nyi, On representing chemical
  environments, Physical Review B 87~(18) (2013) 184115.

\bibitem{bartok2010gaussian}
A.~P. Bart{\'o}k, M.~C. Payne, R.~Kondor, G.~Cs{\'a}nyi, Gaussian approximation
  potentials: The accuracy of quantum mechanics, without the electrons,
  Physical review letters 104~(13) (2010) 136403.

\bibitem{shapeev2016moment}
A.~V. Shapeev, Moment tensor potentials: A class of systematically improvable
  interatomic potentials, Multiscale Modeling \& Simulation 14~(3) (2016)
  1153--1173.

\bibitem{settles2009active}
B.~Settles, Active learning literature survey (2009).

\bibitem{li2024local}
R.~Li, C.~Zhou, A.~Singh, Y.~Pei, G.~Henkelman, L.~Li, Local-environment-guided
  selection of atomic structures for the development of machine-learning
  potentials, The Journal of Chemical Physics 160~(7) (2024).

\bibitem{jinnouchi2020fly}
R.~Jinnouchi, K.~Miwa, F.~Karsai, G.~Kresse, R.~Asahi, On-the-fly active
  learning of interatomic potentials for large-scale atomistic simulations, The
  Journal of Physical Chemistry Letters 11~(17) (2020) 6946--6955.

\bibitem{verdi2021thermal}
C.~Verdi, F.~Karsai, P.~Liu, R.~Jinnouchi, G.~Kresse, Thermal transport and
  phase transitions of zirconia by on-the-fly machine-learned interatomic
  potentials, npj Computational Materials 7~(1) (2021) 156.

\bibitem{podryabinkin2017active}
E.~V. Podryabinkin, A.~V. Shapeev, Active learning of linearly parametrized
  interatomic potentials, Computational Materials Science 140 (2017) 171--180.

\bibitem{zhu2023fast}
A.~Zhu, S.~Batzner, A.~Musaelian, B.~Kozinsky, Fast uncertainty estimates in
  deep learning interatomic potentials, The Journal of Chemical Physics
  158~(16) (2023).

\bibitem{huan2017universal}
T.~D. Huan, R.~Batra, J.~Chapman, S.~Krishnan, L.~Chen, R.~Ramprasad, A
  universal strategy for the creation of machine learning-based atomistic force
  fields, NPJ Computational Materials 3~(1) (2017) 37.

\bibitem{frederiksen2004bayesian}
S.~L. Frederiksen, K.~W. Jacobsen, K.~S. Brown, J.~P. Sethna, Bayesian ensemble
  approach to error estimation of interatomic potentials, Physical review
  letters 93~(16) (2004) 165501.

\bibitem{behler2014representing}
J.~Behler, Representing potential energy surfaces by high-dimensional neural
  network potentials, Journal of Physics: Condensed Matter 26~(18) (2014)
  183001.

\bibitem{smith2017ani}
J.~S. Smith, O.~Isayev, A.~E. Roitberg, Ani-1: an extensible neural network
  potential with dft accuracy at force field computational cost, Chemical
  science 8~(4) (2017) 3192--3203.

\bibitem{jinnouchi2019fly}
R.~Jinnouchi, F.~Karsai, G.~Kresse, On-the-fly machine learning force field
  generation: Application to melting points, Physical Review B 100~(1) (2019)
  014105.

\bibitem{goreinov2010find}
S.~A. Goreinov, I.~V. Oseledets, D.~V. Savostyanov, E.~E. Tyrtyshnikov, N.~L.
  Zamarashkin, How to find a good submatrix, in: Matrix Methods: Theory,
  Algorithms And Applications: Dedicated to the Memory of Gene Golub, World
  Scientific, 2010, pp. 247--256.

\bibitem{novikov2020mlip}
I.~S. Novikov, K.~Gubaev, E.~V. Podryabinkin, A.~V. Shapeev, The mlip package:
  moment tensor potentials with mpi and active learning, Machine Learning:
  Science and Technology 2~(2) (2020) 025002.

\bibitem{podryabinkin2023mlip}
E.~Podryabinkin, K.~Garifullin, A.~Shapeev, I.~Novikov, Mlip-3: Active learning
  on atomic environments with moment tensor potentials, arXiv preprint
  arXiv:2304.13144 (2023).

\bibitem{grigorev2021synergistic}
P.~Grigorev, A.~M. Goryaeva, M.-C. Marinica, J.~R. Kermode, T.~D. Swinburne,
  Synergistic coupling in ab initio-machine learning simulations of
  dislocations, arXiv preprint arXiv:2111.11262 (2021).

\bibitem{pickard2022ephemeral}
C.~J. Pickard, Ephemeral data derived potentials for random structure search,
  Physical Review B 106~(1) (2022) 014102.

\bibitem{PhysRevB.107.125160}
S.~Pozdnyakov, A.~R. Oganov, E.~Mazhnik, A.~Mazitov, I.~Kruglov,
  \href{https://link.aps.org/doi/10.1103/PhysRevB.107.125160}{Fast general two-
  and three-body interatomic potential}, Phys. Rev. B 107 (2023) 125160.
\newblock \href {https://doi.org/10.1103/PhysRevB.107.125160}
  {\path{doi:10.1103/PhysRevB.107.125160}}.
\newline\urlprefix\url{https://link.aps.org/doi/10.1103/PhysRevB.107.125160}

\bibitem{lyakhov2013new}
A.~O. Lyakhov, A.~R. Oganov, H.~T. Stokes, Q.~Zhu, New developments in
  evolutionary structure prediction algorithm uspex, Computer Physics
  Communications 184~(4) (2013) 1172--1182.

\bibitem{poul2023systematic}
M.~Poul, L.~Huber, E.~Bitzek, J.~Neugebauer, Systematic atomic structure
  datasets for machine learning potentials: Application to defects in
  magnesium, Physical Review B 107~(10) (2023) 104103.

\bibitem{avery2017randspg}
P.~Avery, E.~Zurek, Randspg: an open-source program for generating atomistic
  crystal structures with specific spacegroups, Computer Physics Communications
  213 (2017) 208--216.

\bibitem{meziere2023accelerating}
J.~A. Meziere, Y.~Luo, Y.~Xia, L.~K. B{\'e}land, M.~R. Daymond, G.~L. Hart,
  Accelerating training of mlips through small-cell training, Journal of
  Materials Research 38~(24) (2023) 5095--5105.

\bibitem{luo2023set}
Y.~Luo, J.~A. Meziere, G.~D. Samolyuk, G.~L. Hart, M.~R. Daymond, L.~K.
  B{\'e}land, A set of moment tensor potentials for zirconium with increasing
  complexity, Journal of Chemical Theory and Computation 19~(19) (2023)
  6848--6856.

\bibitem{sun2024interatomic}
H.~Sun, C.~Maxwell, E.~Torres, L.~K. B{\'e}land, Interatomic potential for
  sodium and chlorine in both neutral and ionic states, Physical Review B
  109~(17) (2024) 174113.

\bibitem{lide2004crc}
D.~R. Lide, CRC handbook of chemistry and physics, Vol.~85, CRC press, 2004.

\bibitem{perdew1996generalized}
J.~P. Perdew, K.~Burke, M.~Ernzerhof, Generalized gradient approximation made
  simple, Physical review letters 77~(18) (1996) 3865.

\bibitem{prandini2018precision}
G.~Prandini, A.~Marrazzo, I.~E. Castelli, N.~Mounet, N.~Marzari,
  \href{https://www.nature.com/articles/s41524-018-0127-2}{Precision and
  efficiency in solid-state pseudopotential calculations}, npj Computational
  Materials 4~(1) (2018) 72,
  \href{http://materialscloud.org/sssp}{http://materialscloud.org/sssp}.
\newblock \href {https://doi.org/10.1038/s41524-018-0127-2}
  {\path{doi:10.1038/s41524-018-0127-2}}.
\newline\urlprefix\url{https://www.nature.com/articles/s41524-018-0127-2}

\bibitem{dal2014pseudopotentials}
A.~Dal~Corso, Pseudopotentials periodic table: From h to pu, Computational
  Materials Science 95 (2014) 337--350.

\bibitem{schneider1978molecular}
T.~Schneider, E.~Stoll, Molecular-dynamics study of a three-dimensional
  one-component model for distortive phase transitions, Physical Review B
  17~(3) (1978) 1302.

\bibitem{hoover1985canonical}
W.~G. Hoover, Canonical dynamics: Equilibrium phase-space distributions,
  Physical review A 31~(3) (1985) 1695.

\bibitem{giannozzi2009quantum}
P.~Giannozzi, S.~Baroni, N.~Bonini, M.~Calandra, R.~Car, C.~Cavazzoni,
  D.~Ceresoli, G.~L. Chiarotti, M.~Cococcioni, I.~Dabo, et~al., Quantum
  espresso: a modular and open-source software project for quantum simulations
  of materials, Journal of physics: Condensed matter 21~(39) (2009) 395502.

\bibitem{giannozzi2017advanced}
P.~Giannozzi, O.~Andreussi, T.~Brumme, O.~Bunau, M.~B. Nardelli, M.~Calandra,
  R.~Car, C.~Cavazzoni, D.~Ceresoli, M.~Cococcioni, et~al., Advanced
  capabilities for materials modelling with quantum espresso, Journal of
  physics: Condensed matter 29~(46) (2017) 465901.

\bibitem{thompson2022lammps}
A.~P. Thompson, H.~M. Aktulga, R.~Berger, D.~S. Bolintineanu, W.~M. Brown,
  P.~S. Crozier, P.~J. in't Veld, A.~Kohlmeyer, S.~G. Moore, T.~D. Nguyen,
  et~al., Lammps-a flexible simulation tool for particle-based materials
  modeling at the atomic, meso, and continuum scales, Computer Physics
  Communications 271 (2022) 108171.

\bibitem{harris2020array}
C.~R. Harris, K.~J. Millman, S.~J. van~der Walt, R.~Gommers, P.~Virtanen,
  D.~Cournapeau, E.~Wieser, J.~Taylor, S.~Berg, N.~J. Smith, R.~Kern, M.~Picus,
  S.~Hoyer, M.~H. van Kerkwijk, M.~Brett, A.~Haldane, J.~F. del R{\'{i}}o,
  M.~Wiebe, P.~Peterson, P.~G{\'{e}}rard-Marchant, K.~Sheppard, T.~Reddy,
  W.~Weckesser, H.~Abbasi, C.~Gohlke, T.~E. Oliphant,
  \href{https://doi.org/10.1038/s41586-020-2649-2}{Array programming with
  {NumPy}}, Nature 585~(7825) (2020) 357--362.
\newblock \href {https://doi.org/10.1038/s41586-020-2649-2}
  {\path{doi:10.1038/s41586-020-2649-2}}.
\newline\urlprefix\url{https://doi.org/10.1038/s41586-020-2649-2}

\bibitem{Hunter:2007}
J.~D. Hunter, Matplotlib: A 2d graphics environment, Computing in Science \&
  Engineering 9~(3) (2007) 90--95.
\newblock \href {https://doi.org/10.1109/MCSE.2007.55}
  {\path{doi:10.1109/MCSE.2007.55}}.

\bibitem{liu2021alpha}
P.~Liu, C.~Verdi, F.~Karsai, G.~Kresse, $\alpha$-$\beta$ phase transition of
  zirconium predicted by on-the-fly machine-learned force field, Physical
  Review Materials 5~(5) (2021) 053804.

\bibitem{ma2019effect}
P.-W. Ma, S.~Dudarev, Effect of stress on vacancy formation and migration in
  body-centered-cubic metals, Physical Review Materials 3~(6) (2019) 063601.

\bibitem{birch1947finite}
F.~Birch, Finite elastic strain of cubic crystals, Physical review 71~(11)
  (1947) 809.

\bibitem{le2001symmetry}
Y.~Le~Page, P.~Saxe, Symmetry-general least-squares extraction of elastic
  coefficients from ab initio total energy calculations, Physical Review B
  63~(17) (2001) 174103.

\bibitem{sholl2022density}
D.~S. Sholl, J.~A. Steckel, Density functional theory: a practical
  introduction, John Wiley \& Sons, 2022.

\bibitem{henkelman2000climbing}
G.~Henkelman, B.~P. Uberuaga, H.~J{\'o}nsson, A climbing image nudged elastic
  band method for finding saddle points and minimum energy paths, The Journal
  of chemical physics 113~(22) (2000) 9901--9904.

\bibitem{greenfield1971x}
A.~Greenfield, J.~Wellendorf, N.~Wiser, X-ray determination of the static
  structure factor of liquid na and k, Physical Review A 4~(4) (1971) 1607.

\bibitem{zhang2012comparison}
Y.~Zhang, E.~J. Maginn, A comparison of methods for melting point calculation
  using molecular dynamics simulations, The Journal of chemical physics
  136~(14) (2012).

\bibitem{miryashkin2023bayesian}
T.~Miryashkin, O.~Klimanova, V.~Ladygin, A.~Shapeev, Bayesian inference of
  composition-dependent phase diagrams, Physical Review B 108~(17) (2023)
  174103.

\bibitem{sengers1986thermodynamic}
J.~Sengers, J.~L. Sengers, Thermodynamic behavior of fluids near the critical
  point, Annual Review of Physical Chemistry 37~(1) (1986) 189--222.

\bibitem{pelissetto2002critical}
A.~Pelissetto, E.~Vicari, Critical phenomena and renormalization-group theory,
  Physics Reports 368~(6) (2002) 549--727.

\bibitem{leonchuk2022nak}
S.~S. Leonchuk, A.~S. Falchevskaya, V.~Nikolaev, V.~V. Vinogradov, Nak alloy:
  underrated liquid metal, Journal of Materials Chemistry A 10~(43) (2022)
  22955--22976.

\bibitem{henninger1966atomic}
E.~Henninger, R.~Buschert, L.~Heaton, Atomic structure and correlation in
  liquid binaries by x-ray and neutron diffraction with application to nak, The
  Journal of Chemical Physics 44~(5) (1966) 1758--1764.

\bibitem{foust1976sodium}
O.~Foust, Sodium-NaK engineering handbook, Vol.~2, ME Sharpe, 1976.

\bibitem{sadigh2012scalable}
B.~Sadigh, P.~Erhart, A.~Stukowski, A.~Caro, E.~Martinez, L.~Zepeda-Ruiz,
  Scalable parallel monte carlo algorithm for atomistic simulations of
  precipitation in alloys, Physical Review B—Condensed Matter and Materials
  Physics 85~(18) (2012) 184203.

\bibitem{stukowski2009visualization}
A.~Stukowski, Visualization and analysis of atomistic simulation data with
  ovito--the open visualization tool, Modelling and simulation in materials
  science and engineering 18~(1) (2009) 015012.

\bibitem{huang2022ab}
Y.~Huang, M.~Widom, M.~C. Gao, Ab initio free energies of liquid metal alloys:
  Application to the phase diagrams of li-na and k-na, Physical Review
  Materials 6~(1) (2022) 013802.

\bibitem{reitz2019simulating}
D.~M. Reitz, E.~Blaisten-Barojas, Simulating the nak eutectic alloy with monte
  carlo and machine learning, Scientific Reports 9~(1) (2019) 704.

\bibitem{reitz2017monte}
D.~Reitz, E.~Blaisten-Barojas, Monte carlo study of the crystalline and
  amorphous nak alloy, Procedia Computer Science 108 (2017) 1215--1221.

\bibitem{nitol2023hybrid}
M.~S. Nitol, K.~Dang, S.~J. Fensin, M.~I. Baskes, D.~E. Dickel, C.~D. Barrett,
  Hybrid interatomic potential for sn, Physical Review Materials 7~(4) (2023)
  043601.

\bibitem{ziegler1985stopping}
J.~F. Ziegler, J.~P. Biersack, The stopping and range of ions in matter, in:
  Treatise on heavy-ion science: volume 6: astrophysics, chemistry, and
  condensed matter, Springer, 1985, pp. 93--129.

\bibitem{byggmastar2019machine}
J.~Byggm{\"a}star, A.~Hamedani, K.~Nordlund, F.~Djurabekova, Machine-learning
  interatomic potential for radiation damage and defects in tungsten, Physical
  Review B 100~(14) (2019) 144105.

\bibitem{wang2019deep}
H.~Wang, X.~Guo, L.~Zhang, H.~Wang, J.~Xue, Deep learning inter-atomic
  potential model for accurate irradiation damage simulations, Applied Physics
  Letters 114~(24) (2019).

\bibitem{wood2017quantum}
M.~A. Wood, A.~P. Thompson, Quantum-accurate molecular dynamics potential for
  tungsten, arXiv preprint arXiv:1702.07042 (2017).

\bibitem{annevelink2023statistical}
E.~Annevelink, V.~Viswanathan, Statistical methods for resolving poor
  uncertainty quantification in machine learning interatomic potentials, arXiv
  preprint arXiv:2308.15653 (2023).

\bibitem{kulichenko2024data}
M.~Kulichenko, B.~Nebgen, N.~Lubbers, J.~S. Smith, K.~Barros, A.~E. Allen,
  A.~Habib, E.~Shinkle, N.~Fedik, Y.~W. Li, et~al., Data generation for machine
  learning interatomic potentials and beyond, Chemical Reviews 124~(24) (2024)
  13681--13714.

\bibitem{batatia2022mace}
I.~Batatia, D.~P. Kovacs, G.~Simm, C.~Ortner, G.~Cs{\'a}nyi, Mace: Higher order
  equivariant message passing neural networks for fast and accurate force
  fields, Advances in neural information processing systems 35 (2022)
  11423--11436.

\bibitem{thompson2015spectral}
A.~P. Thompson, L.~P. Swiler, C.~R. Trott, S.~M. Foiles, G.~J. Tucker, Spectral
  neighbor analysis method for automated generation of quantum-accurate
  interatomic potentials, Journal of Computational Physics 285 (2015) 316--330.

\bibitem{drautz2019atomic}
R.~Drautz, Atomic cluster expansion for accurate and transferable interatomic
  potentials, Physical Review B 99~(1) (2019) 014104.

\bibitem{earl2005parallel}
D.~J. Earl, M.~W. Deem, Parallel tempering: Theory, applications, and new
  perspectives, Physical Chemistry Chemical Physics 7~(23) (2005) 3910--3916.

\bibitem{swendsen1986replica}
R.~H. Swendsen, J.-S. Wang, Replica monte carlo simulation of spin-glasses,
  Physical review letters 57~(21) (1986) 2607.

\end{thebibliography}






\end{document}